\documentclass[9pt,twocolumn,twoside]{osajnl}

\usepackage{subfig,graphicx,ulem}
\usepackage[english]{babel}
\usepackage{physics}
\journal{josab}
\setboolean{shortarticle}{true}

\DeclareUnicodeCharacter{03BC}{$\mu$}

\title{Fiber-based photon pair generation: a tutorial}

\author[1]{Karina Garay-Palmett}
\author[2]{Dong Beom Kim}
\author[2]{Yujie Zhang}
\author[3]{Francisco A. Dominguez-Serna}
\author[2]{Virginia O. Lorenz}
\author[4*]{Alfred B. U'Ren}

\affil[1]{Departamento de \'Optica - Centro de Investigaci\'on Cient\'ifica y de Educaci\'on Superior de \\ Ensenada, B.C., 22860, M\'exico}
\affil[2]{Department of Physics, University of Illinois at Urbana-Champaign, Urbana, Illinois 61801, USA}
\affil[3]{ C\'atedras Conacyt - Centro de Investigaci\'on Cient\'ifica y de Educaci\'on Superior de \\ Ensenada, B.C., 22860, M\'exico}
\affil[4]{ Instituto de Ciencias Nucleares, Universidad Nacional Aut\'onoma de M\'exico, A.P. 70-543, 04510 Ciudad de M\'exico, M\'exico}

\affil[*]{alfred.uren@correo.nucleares.unam.mx, kgaray@cicese.edu.mx}

\begin{abstract}
The purpose of this tutorial paper is to present a broad overview of photon-pair generation through the spontaneous four wave mixing (SFWM) process in optical fibers. Progress in optical fiber technology means that today we have at our disposal a wide variety of types of fiber, which together with the fact that SFWM uses two pump fields, implies a truly remarkable versatility in the resulting possible photon-pair properties.    We discuss how the interplay of the frequency, transverse mode, and polarization degrees of freedom, the first linked to the latter two through fiber dispersion, leads to interesting entanglement properties both in individual degrees of freedom and also permitting hybrid and hyper entanglement in combinations of degrees of freedom.    This tutorial covers methods for photon pair factorability, frequency tunability, and SFWM bandwidth control, the effect of frequency non-degenerate and counter-propagating pumps,  as well methods for characterizing photon pairs generated in optical fibers.
\end{abstract}

\begin{document}
\maketitle

\section{Introduction}

Photon pair sources are at the heart of the remarkable progress made  over the past two decades in the field of  optical quantum information processing \cite{Zoller2005}. There are two main experimental avenues for photon pair generation: the spontaneous parametric downconversion (SPDC) process based on second-order nonlinear crystals \cite{Kim:05}, and the spontaneous four wave mixing (SFWM) process based on third-order nonlinearities \cite{Fiorentino2002}. Typically, SFWM implementations rely on the one hand on waveguides or  micro-resonators (including integrated optical designs) \cite{Helt:10}, and on the other hand on optical fibers \cite{Li:04,Rarity:05,Fan:05,Fulconis_2007,Cohen2009,PETROV2019}. Compared to SPDC, SFWM is advantageous because the two pumps on which the process is based afford a greater scope for source tailoring to particular needs \cite{Garay-Palmett:07}. Indeed, as will be discussed below, making the two pumps non-degenerate spectrally, spatially, and/or in terms of propagation direction leads to useful avenues for photon-pair state engineering.  Amongst SFWM sources, those based on optical fiber exhibit a tremendous versatility in terms of the resulting photon pair properties because of the enormous variety of fibers available, including for example photonic crystal \cite{XiaoyingLi2018}, tapered \cite{Shukhin2020}, birefringent \cite{Smith:09}, gas-filled hollow-core \cite{Cordier2020}, single \cite{Rarity:05}- and few-mode \cite{Guo:19}, amongst others.


Fiber-based SFWM implementations exhibit a number of distinct advantages in terms of the attainable brightness over SPDC sources, including i) a quadratic dependence of the emitted flux on the pump power, in contrast to the linear dependence of SPDC, ii)  the possibility of an essentially unlimited length of the nonlinear element, leading to larger interaction lengths, iii) an effective non-linearity which can be scaled by reducing the transverse mode area, and iv) since photon pairs are generated in fiber-modes, coupling into additional optical fibers in a given experiment (beyond the fiber used as the source) can be attained with very high efficiency.  We discuss these advantages in more detail in section 4.

The purpose of this paper is to provide an ample overview, in tutorial form, of photon-pair generation in optical fibers.   This overview includes in sections 2-6 a description of the two-photon state, with a discussion of the different combinations of transverse and polarization modes (which become correlated with the optical frequency of emission through fiber dispersion), and the resulting various phasematching configurations.  In sections 7-10 we describe  methods for factorable photon-pair / pure heralded single photon  generation, methods for spectrally tuning the two-photon state, and SFWM bandwidth control (including generation of ultra-narrow and ultra-broadband photon pairs).   In sections 11-13  we describe methods to control the transverse mode, spontaneous Raman scattering as a noise mechanism, as well as SFWM source variations including the use of a dual pump (spectrally non-degenerate as well as counterpropagating) and entanglement generation in  energy-time, polarization \cite{Lee2022}, frequency-transverse mode, frequency–polarization, and discrete frequency.  In section 14 we cover the various techniques that have been developed for photon pair characterization, in both the spectral and spatial domains.

We hope that this tutorial paper will constitute a comprehensive review of the field of photon pair generation based on optical fibers, and as such can serve as an adequate starting point for a reader interested in embarking on new research in this topic.

\section{The spontaneous four-wave mixing process}
The spontaneous four-wave mixing (SFWM) process is a parametric third-order nonlinear interaction in which photon pairs, conventionally called signal and idler photons, are created upon the annihilation of two pump photons, mediated by vacuum fluctuations. The process is constrained to the fulfillment of energy conservation and the phasematching condition, which in general, for an optical fiber, take the form: $\Delta\omega=\omega_{p1}+\omega_{p2}-\omega_s-\omega_i=0$, and $\Delta k=k_{p1}(\omega_{p1})+k_{p2}(\omega_{p2})-k_{s}(\omega_{s})-k_{i}(\omega_{i})-(\gamma_1 P_1+\gamma_2 P_2)=0$, respectively, where $\omega_{\mu}$ ($\mu=p1,p2,s,i$) is the electric field frequency, $k_{\mu}(\omega_{\mu})$ is the propagation constant, $\gamma_{\nu}$ ($\nu=1,2$) is the effective nonlinear coefficient related to the self/cross phase modulation effects, and $P_{\nu}$ are the peak pump powers \cite{Garay-Palmett:07}. 

The photon pairs generated by SFWM in general exhibit correlations in continuous-variable degrees of freedom, such as frequency and transverse momentum. However, in media characterized by a single transverse mode (single-mode optical fibers or rectangular waveguides), trasnverse momentum correlations are suppressed.  In few- or multi-mode fibers or waveguides, correlations in discrete degrees of freedom, such as polarization, temporal modes, transverse waveguide modes, and orbital angular momentum, can still emerge, depending on waveguide dispersion properties and pump field characteristics \cite{Cruz-Delgado2016,Daniel2021}. The SFWM process in specialty or conventional optical fibers has led to the proposal or demonstration of entangled two-photon states in several degrees of freedom, e.g., polarization  \cite{Zhu:16}, energy-time/polarization  \cite{Dong2015}, discrete frequency \cite{Zhou2014}, spatial entanglement \cite{Ekici2020}, among others. 

\section{The SFWM two-photon state}
The SFWM process can occur in different polarization and transverse mode combinations amongst the four waves involved.   For the treatment below, we will assume that the fiber is birefringent, leading to the appearance of slow and fast principal axes (labelled $x$ and $y$, respectively), and that it supports more than one transverse mode.    From the third-order nonlinear polarization of the medium, and following a standard perturbative approach \cite{mandel1995}, it can be shown that the SFWM two-photon state can be written in terms of a coherent superposition of the contributions from the different polarization and transverse mode combinations. The general two-photon state is then given by $\ket{\Psi}=\ket{\mbox{vac}}+\eta\ket{\Psi}_2$, with $\ket{\mbox{vac}}$ representing the vacuum, $\eta$  a global constant related to the conversion efficiency, and $\ket{\Psi}_2$ the two-photon component of the state given as 

\begin{equation}
	\begin{aligned}
             \ket{\Psi}_2 &= \sum_{m=1}^M\sum_{n=1}^6\kappa_{mn}\sqrt{p_{mn1}p_{mn2}}\int\!\! d\omega_s\!\!\int \!\!d\omega_i\, G_{mn}(\omega_s,\omega_i)\\ &\quad\times\hat{a}_s^\dagger (\omega_s; q_m; \phi_n) \hat{a}_i^\dagger (\omega_i; r_m; \varphi_n)\ket{\mbox{vac}};
	\end{aligned}
	\label{eqn:state}
\end{equation} 

\noindent where the index $m$ represents each of the transverse mode combinations amongst the four waves involved (pump 1, pump 2, signal, and idler), an $n$ represents each of the six different polarization combinations allowed in an isotropic medium, such as fused silica (see table \ref{tab:process}) \cite{Daniel2021, Agrawal2008}. The term $\kappa_{mn}$ is a coefficient weighing each SFWM interaction, which depends on the transverse mode overlap between the four waves (pump 1, pump 2, signal, and idler), $p_{mn\nu}$ ($\nu=1,2$) represents the average pump power for each of the two pump modes, $G_{mn}(\omega_s,\omega_i)$ is the joint spectral amplitude (JSA) function, and $\hat{a}_{\mu}^\dagger (\omega_{\mu}; q_m; \phi_n) $ ($\mu=s,i$) is the creation operator of photons at frequency $\omega_{\mu}$, transverse mode $q_m$ or $r_m$, and polarization $\phi_n$ or $\varphi_n$.

\begin{table}[t]
\centering
\caption{\bf List of the SFWM processes related to different polarization combinations of the participating electric fields in single-transverse-mode fiber.}
\begin{tabular}{ccccc}
\hline 
process &$p_1$& $p_2$ & s \,($\lambda_s\textit{\textcolor{blue}{<}}\lambda_p$) & i \,($\lambda_p\textit{\textcolor{blue}{<}}\lambda_i$)\\
\hline
1 &x&x & x& x\\
2 &y&y & y&y\\
3 &x&y& x& y\\
4 &x&y& y& x\\
5 &x&x& y& y\\
6 &y&y& x& x\\
\hline
\end{tabular}
  \label{tab:process}
\end{table}

For fixed values of $m$ and $n$, Eq. \ref{eqn:state} is reduced to a single term. This would be the case, for example, for an SFWM interaction in which all fields involved propagate in the fundamental transverse mode and are co-polarized along one of the two axes. In this case, the JSA is given by

\begin{equation}
	\begin{aligned}
      F(\omega_s,\omega_i)=\int\! d\omega\,\alpha(\omega)\alpha(\omega_s+\omega_i-\omega)\mbox{sinc}\left[\frac{L}{2}\Delta k\right]\mbox{exp}\left[i\frac{L}{2}\Delta k\right],
	\end{aligned}
	\label{eqn:jsa}
\end{equation} 

\noindent where $\Delta k$ was defined in the previous section, $L$ is the length of the nonlinear medium, and $\alpha(\omega)$ is the pump spectral envelope function, assumed to be normalized so that $\int\! d\omega\, |\alpha(\omega)|^2=1$. For pump pulses with a Gaussian spectral envelope, expanding the propagation constant as a Taylor series around the central frequencies for which phasematching is fulfilled up to first-order, the JSA can be expressed as \cite{Garay-Palmett:07}

\begin{equation}
	\begin{aligned}
         F_{lin}(\nu_s,\nu_i)=\alpha(\nu_s+\nu_i)\phi(\nu_s,\nu_i),
	\end{aligned}
	\label{eqn:jsalin}
\end{equation} 

\noindent where $lin$ refers to the linear approximation of the phase-mismatch function, $\alpha(\nu_s+\nu_i)$ is the pump spectral envelope function, $\phi(\nu_s,\nu_i)$ is the phasematching function, and $\nu_{\mu}=\omega_{\mu}-\omega_{\mu 0}$, with $\omega_{\mu 0}$ the central emission frequency.

  \begin{figure*}[t!]
  \begin{center}
  		\includegraphics[width=0.7\linewidth]{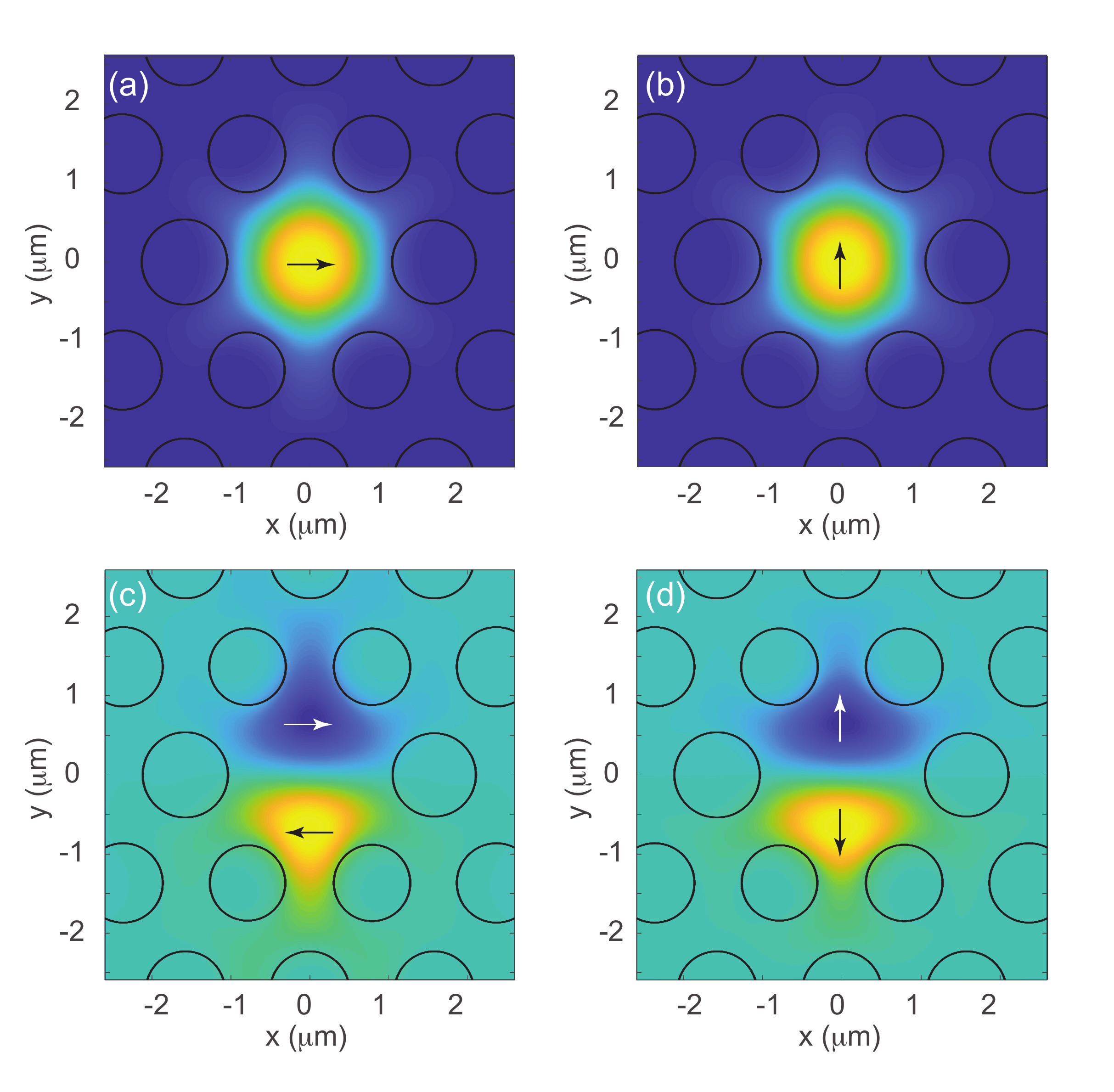}
  	\caption{\footnotesize{Transverse modes structure of the fiber under study.  (a) $HE_{11}^x$, (b) $HE_{11}^x$, (c) $TE_{01}$ and (d) $TM_{01}$.}}
  	\label{fig:modes}
  	\end{center}
  \end{figure*}

    \begin{figure*}[t!]
  \begin{center}
  		\includegraphics[width=0.7\linewidth]{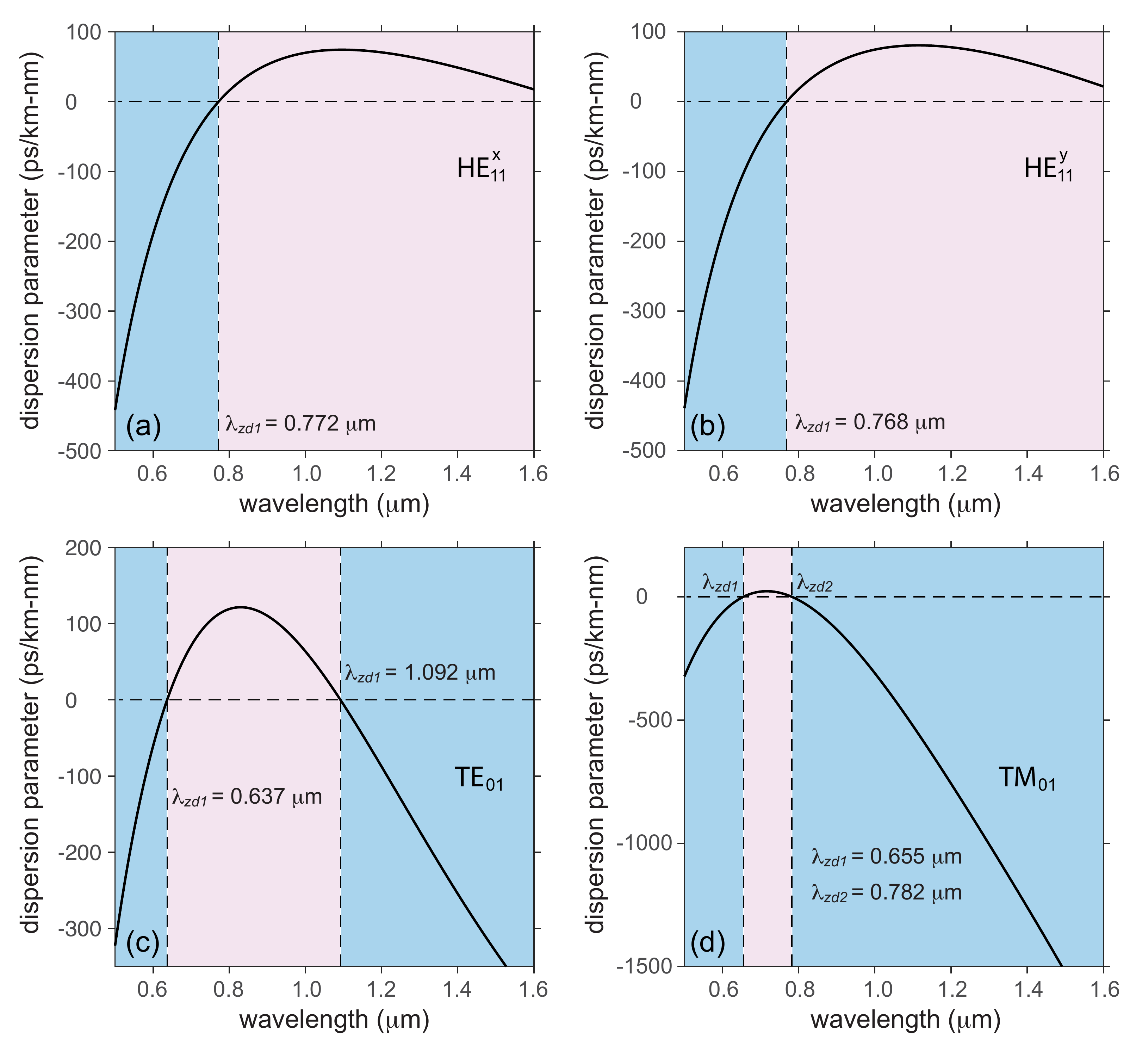}
  	\caption{\footnotesize{Fiber dispersion parameter for the transverse modes: (a) $HE_{11}^x$, (b) $HE_{11}^x$, (c) $TE_{01}$ and (d) $TM_{01}$, shown in figure \ref{fig:modes}. Blue zones correspond to normal dispersion, and pink zones to anomalous dispersion.}}
  	\label{fig:dispersionParameter}
  	\end{center}
  \end{figure*}

\section{Brightness of SFWM and SPDC sources}\label{sec:BrCmp}

The brightness of SFWM sources can be described by the emitted photon-pair flux $R_{SFWM}$. For degenerate pumps, $R_{SFWM}$ has been shown to be proportional to the square of the average pump power $p$, the pump bandwidth $\sigma$, the fiber length $L$, and the square of the effective nonlinear coefficient $\gamma$ \cite{Garay2010,Chen2005,Lin2007},

\begin{equation} \label{eq:Rsfwm}
    R_{SFWM}\propto p^2 \sigma L \gamma^2.
\end{equation}

For non-degenerate pumps, on the other hand, $p^2$ is replaced by $p_1p_2$, the product of two pump powers involved. In addition, there is a limit to the maximum interaction length, $L_{max}$, due to the longitudinal walk-off between the two pumps having different group velocities. Thus, for $L\leq L_{max}$, the photon-pair flux increases linearly, and for $L>L_{max}$, the emission ceases (see section~\ref{SFWMvariation}B for more details).

In this respect, fiber-based SFWM implementations exhibit a number of distinct advantages over SPDC sources in terms of the attainable brightness \cite{Garay2010,zhang2012heralded} as outlined below.

First, the brightness of SFWM sources scales better with incident pump power than that of the SPDC sources. As discussed earlier, while the SFWM brightness scales either quadratically or as the product of two pump powers, the SPDC brightness scales linearly with the pump power. Hence, with sufficiently large pump power, the brightness of a given SFWM source can exceed that of an SPDC source. This is manifested in \cite{Smith:09,Soller2011,goldschmidt2008spectrally,PETROV2019} for SFWM, where the raw rates of average photon pairs detected ($\sim10^4$ pairs/s, or $\sim10^4$ pairs/s/$\mathrm{mW^2}$ to illustrate the quadratic scaling more appropriately with the assumed $\sim$1 mW pump power) are similar as those of SPDC sources ($\sim10^4$ pairs/s, or $\sim10^3$ pairs/s/mW, with a $\sim$10 mW pump power)~\cite{chen2017efficient,pickston2021optimised}, even with an order of magnitude lower pump power and in the absence of superconducting nano-wire single photon detectors (SNSPDs).

Second, the accessible lengths of fiber-based SFWM sources are substantially longer than those of SPDC sources -- fabrication and mounting considerations limit the corresponding SPDC crystal lengths to the order of few cm. Note, however, an increase in source brightness from scaling the fiber length accompanies changes in spectral photon-pair characteristics since phasematching properties are highly dependent on the fiber length. Similar to the $L_{max}$ constraint for non-degenerate pumps previously mentioned, degenerate pumps also require a slight caveat when intended to be utilized with a long fiber: for degenerate pumps, length-dependent chirp may temporally broaden the pump pulses, thus changing which frequency components can overlap temporally \cite{Fang2014}. This, in turn, results in modifying the photon-pair spectral characteristics and may constrain the total photon-pair flux.

Third, the transverse confinement of a fiber with a small core radius can significantly boost the emission rate. The effective nonlinearity $\gamma$ in Eq.~\ref{eq:Rsfwm} of fiber-based SFWM sources is inversely proportional to the transverse mode area $A$, resulting in a $A^{-2}$ dependence on the emitted flux. Note, however, that this advantage is not exclusive to fiber-based SFWM, and  waveguide-based SPDC sources will have similar behavior. Nevertheless, the vast variety of fiber types explored as SFWM sources (see section~\ref{SFWMvariation}A) and their commercial availability with various core sizes allow for more flexible and approachable exploitation of this $A^{-2}$ scaling benefit.

Lastly, fiber-based SFWM sources naturally facilitate coupling into optical fibers (that may follow the source fiber) as the created photon pairs are already in well-defined fiber spatial modes \cite{Soller2011}. On the contrary, efficient fiber coupling is often challenging for bulk-crystal SPDC sources \cite{schwaller2022optimizing,Vicent_2010,ling2008absolute}. Even in the case of waveguide-based SPDC, coupling losses can be significant, since the fiber and waveguide modes may exhibit poor mode matching \cite{Zhao20,Jung14}. Importantly, fiber-based SFWM sources permit straightforward integration with fiber optic networks \cite{Medic:10} and full access to the current optical fiber technology.

We now briefly consider additional metrics, for which SFWM sources have comparable quantitative performance to their SPDC counterparts: heralded single-photon purity, heralding efficiency, and entanglement fidelity (see section~\ref{sec:characterization} and \cite{anwar2021entangled,nielsen_chuang_2010} for definitions of these metrics). When factorable photon-pair sources are used as heralded single-photon sources, both SFWM~\cite{Liu:16,Cohen2009,Soller2011,Zhang2019} and SPDC~\cite{mosley2008heralded,kaneda2016heralded,chen2017efficient} sources have demonstrated purities that are close to or above $90\%$. In terms of heralding efficiency, an important metric not only to assess the performance of a heralded source but also for applications in quantum information science~\cite{zhang2012heralded,datta2011quantum,vertesi2010closing}, SFWM~\cite{Soller2011} and SPDC~\cite{kaneda2016heralded,pickston2021optimised} both have shown $>80\%$ normalized heralding efficiencies. Here, the normalized heralding efficiency, sometimes called collection efficiency, refers to a heralding efficiency normalized with the detection efficiency~\cite{anwar2021entangled,kaneda2016heralded,klyshko1980use} and is often found in the literature to emphasize the detection-independent properties intrinsic to the source \cite{goldschmidt2008spectrally,Soller2011,pickston2021optimised}. Analogously, the entanglement fidelity, often a Bell-state fidelity (fidelity with respect to the target Bell state), is similar in both SFWM and SPDC sources. Bell-state fidelities of $>95\%$ have been demonstrated with SFWM~\cite{Fang2014,fang2016multidimensional,Li09} and SPDC sources~\cite{anwar2021entangled,evans2010bright,lohrmann2018high,steinlechner2014efficient,kaiser2014polarization}. It is also worth noting that both SFWM and SPDC sources have been reported (just to name a few here) to produce photon pairs in near-infrared \cite{Fang2014,meyer2013generating,Cohen2009,Smith:09,goldschmidt2008spectrally,Soller2011}, \cite{mosley2008heralded,schwaller2022optimizing} and telecommunications \cite{Rarity:05,XiaoyingLi2018,Liu:16}, \cite{pickston2021optimised,chen2017efficient,kaneda2016heralded} wavelengths.

\section{Phasematching configurations}\label{PM}

  \begin{figure*}[t!]
  \begin{center}
  		\includegraphics[width=0.8\linewidth]{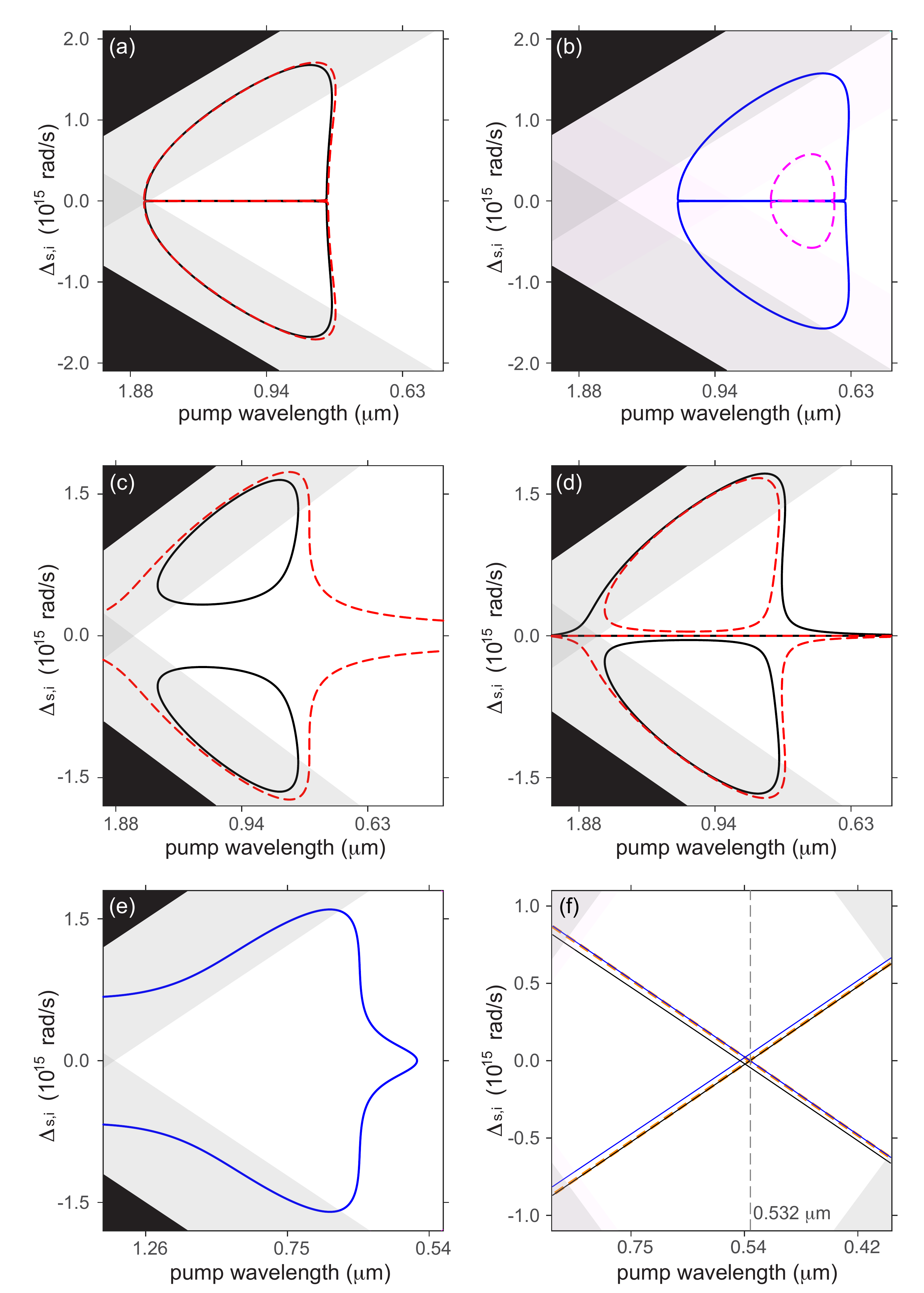}
  	\caption{\footnotesize{Phasematching contours for SFWM processes characterized by the following combinations of transverse modes for the four waves involved (pump 1, pump 2, signal, idler).  (a) $HE_{11}^x + HE_{11}^x\rightarrow HE_{11}^x + HE_{11}^x $ (solid black), and $HE_{11}^y + HE_{11}^y\rightarrow HE_{11}^y + HE_{11}^y $ (dashed red). (b) $TE_{01} + TE_{01}\rightarrow TE_{01} + TE_{01} $ (solid blue), and  $TM_{01} + TM_{01}\rightarrow TM_{01} + TM_{01} $ (dashed magenta) (c) $HE_{11}^y + HE_{11}^y\rightarrow HE_{11}^x + HE_{11}^x $ (solid black), and $HE_{11}^x + HE_{11}^x\rightarrow HE_{11}^y + HE_{11}^y $ (dashed red). (d) $HE_{11}^x + HE_{11}^y\rightarrow HE_{11}^x + HE_{11}^y $ (solid black), and $HE_{11}^x + HE_{11}^y\rightarrow HE_{11}^y + HE_{11}^x $ (dashed red). (e) $TM_{01} + TM_{01}\rightarrow TE_{01} + TE_{01} $ (solid blue). (f) $HE_{11}^x - HE_{11}^x\rightarrow -HE_{11}^x + HE_{11}^x $ (dashed yellow); $+HE_{11}^y$ - $TM_{01}$  $\rightarrow$ $-TE_{01}$+$HE_{11}^x$ and $+HE_{11}^y$ - $TM_{01}$  $\rightarrow$ $+TE_{01}$-$HE_{11}^x$ (solid blue); $+HE_{11}^y$ - $TM_{01}$  $\rightarrow$ $-HE_{11}^x$ + $TE_{01}$ and $+HE_{11}^y$ - $TM_{01}$  $\rightarrow$ $+HE_{11}^x$ + $-TE_{01}$ (solid black).  Note: +/- signs indicate forward/backward propagation.}}
  	\label{fig:PMContours}
  	\end{center}
  \end{figure*} 

Optical fibers constitute a versatile system for photon-pair generation through the SFWM process. It is important to remark that optical fibers can lead to phasematching for different combinations of transverse and polarization modes, which in turn become correlated to the emission frequency through fiber dispersion.  This opens up the possibility for the preparation of two-photon states with entanglement in one of these degrees of freedom individually, or alternatively hybrid- or hyper-entanglement involving more than one of these degrees of freedom.   Examples of the above include polarization entanglement \cite{Zhu:16}, frequency entanglement \cite{Zhou2014}, transverse-mode entanglement \cite{Ekici2020}, spatial-frequency correlations \cite{Cruz-Delgado2016}, polarization-frequency correlations \cite{Daniel2021}, spatial-polarization-frequency correlations \cite{majchrowska2022multiple}, as well hyper-entanglement in polarization-frequency \cite{Dong2015}.

The above discussion leads us to explore different SFWM phasematching configurations. To this end, we consider as reference a commercial, polarization-maintaining photonic crystal fiber (PCF) --NL-PM-750 from NKT Photonics-- which is single-mode over a wide spectral range.  We note that scaling the transverse PCF structure (e.g. by pulling under the action of heat), leads to the ability to tune the resulting photon-pair properties, for example the emission frequency \cite{Ortiz_Ricardo_2017,Kihara1996}.

Let us assume a PCF with its transverse dimensions increased by $10\%$ with respect to the reference fiber, which allows for the propagation of the fundamental transverse modes in its two orthogonal polarizations $HE_{11}^x$ and $HE_{11}^y$, as well as the higher transverse modes $TE_{01}$ and $TM_{01}$, which are orthogonally polarized (along $x$ and $y$, respectively) with the same parity.  Unless otherwise noted, this fiber is assumed in all of the simulations presented in sections 2-11.  In figure \ref{fig:modes} we show the transverse intensity distributions $\vert $f(x,y)$ \vert^2$ of the mentioned modes, superimposed with the PCF structure.  
Note also that we limit our discussion below to these four modes, thus restricting the number of SFWM processes resulting from transverse / polarization mode combinations.

It is instructive to plot the fiber dispersion parameter ($D=-\frac{\lambda}{c}\frac{d^2n}{d\lambda^2}$) vs wavelength, for each of these four modes, as is done in figure \ref{fig:dispersionParameter}.
Note that within the spectral region shown the fundamental mode in its two polarizations exhibits one so-called zero-dispersion wavelength (ZDW) for which $D=0$, at which an optical pulse travels with zero quadratic dispersion; see panels (a) and (b) \cite{Agrawal2008}.   We have shown in blue [pink]  areas the normal dispersion regime ($\lambda<\lambda_{zd}$)  [anomalous dispersion regime ($\lambda>\lambda_{zd}$)]. Interestingly, the higher transverse modes (TE$_{01}$ and TM$_{01}$) exhibit two ZDWs, see panels (c) and (d) \cite{Saitoh_2006}. 
As discussed below, the ZDWs are intimately related to the phasematching properties for single-transverse mode and co-polarized degenerate-pump SFWM interactions; while for a single ZDW phasematching occurs for pump wavelengths around $ \lambda_{zd}$, for two ZDWs $\lambda_{ZD1}$ and $\lambda_{ZD2}$ phasematched pump wavelengths are contained by the two ZDWs, i.e. $\lambda_{ZD1}<\lambda<\lambda_{ZD2}$.

Figure \ref{fig:PMContours} shows the phasematching diagrams, i.e. composed of solutions to $\Delta k=0$  where the vertical axis represents the SFWM frequencies, as a detuning from the pump $\Delta_{s,i}=\omega_{s,i}-\omega_p$, and the horizontal axis represents the pump frequency $\omega_p$, for different combinations of the modes depicted in figure \ref{fig:modes}.  In these diagrams, the top half ($\Delta>0$) corresponds to the idler photon whereas the bottom half ($\Delta<0$) corresponds to the signal; note that energy conservation imposes a symmetry between these two halves.   Black-shaded areas correspond to non-physical regions in the sense that one or both SFWM frequencies would have to be negative to fulfill energy conservation.    Note that these phasematching curves are based on the numerical evaluation of the fiber dispersion relations employing a  numerical finite-difference time-domain vectorial mode solver; note that gray-shaded areas lie outside the spectral region of the numerical evaluation, so that these portions of the contours can be considered as extrapolations and may depart from the actual observed behavior. 
We have used the convention that solid lines in black (blue) correspond to photons emitted in the $HE_{11}^x$ ($TE_{01}$) mode,  while dashed lines in red  (magenta) represent emitted signals traveling in the  $HE_{11}^y$ ($TM_{01}$) mode.

\begin{figure}[t!]
  \begin{center}
  		\includegraphics[width=1.0\linewidth]{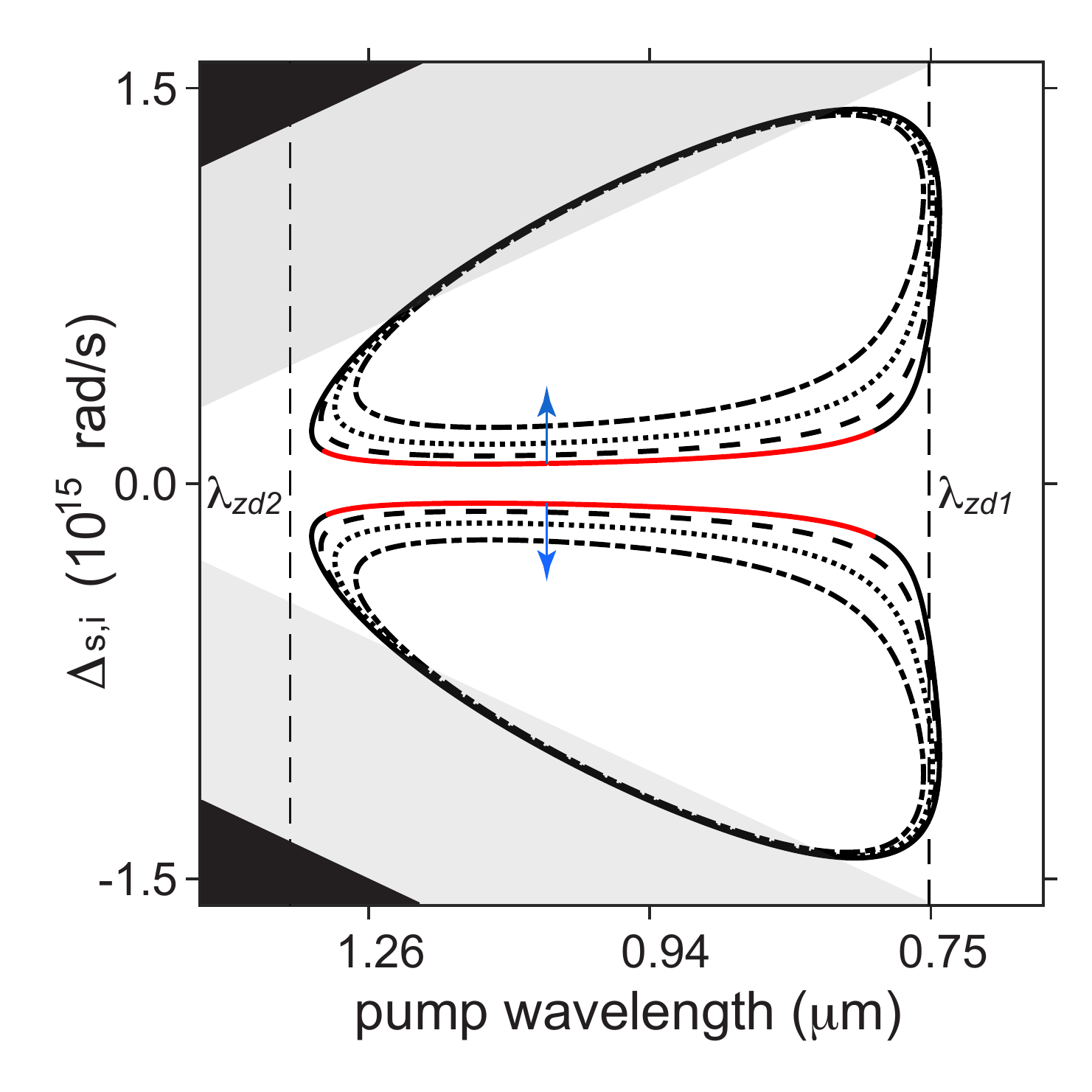}
  	\caption{Self/cross phase nonlinear shifting in SFWM signals for the NL-PM-750 fiber. The solid contour corresponds to a nonlinear shifting of $240\times10^{-6}\,\mu \mbox{m}^{-1}$; the red(black) line represents the inner(outer) solution to $\Delta k =0$. Dashed black lines correspond to nonlinear shiftings of $480\times10^{-6}\,\mu \mbox{m}^{-1}$,  $960\times10^{-6}\,\mu\mbox{m}^{-1}$, and $1920\times10^{-6}\,\mu\mbox{m}^{-1}$, as indicated by the blue arrows. Note that the two zero-dispersion wavelengths have been indicated.}
	  	\label{fig:inner_outer}
  	\end{center}
  \end{figure}

Let us first consider cases in which all four waves propagate in the same mode, whether the fundamental modes $HE_{11}^x$ and $HE_{11}^y$ (panel a) or the non-fundamental modes $TE_{01}$  and $TM_{01}$ (panel b),
corresponding to processes 1 and 2 in table \ref{tab:process}. In these cases, if the optical fiber exhibits two points of zero dispersion ($\lambda_{ZD1}$, $\lambda_{ZD2}$) in the spectral region of interest, the $\Delta k=0$ contours take the form of two closed loops, constrained by $\lambda_{ZD1}$ and $\lambda_{ZD2}$, as can be seen in panels a) and b). The $\Delta k=0$ contour's splitting into two loops results from the modulation instability of the pump field and is controlled by the pump peak power, leading to a nonlinear shift from the trivial solution (see figure \ref{fig:inner_outer}) \cite{Tai1986, FATOME2006,Erkintalo2011}. In contrast, if the optical fiber is characterized by just one zero-dispersion wavelength, the $\Delta k=0$ contour consists of nearly parallel lines for pump wavelengths $\lambda_p<\lambda_{ZD}$, which open up sharply for $\lambda_p>\lambda_{ZD}$.


Let us now explore the phasematching properties of interactions in which the four participating fields travel in the same transverse mode but involve the two polarizations, corresponding to processes 3-6 in table \ref{tab:process}. 
In figure \ref{fig:PMContours} (c) we show the phasematching contours for processes $5$ and $6$, which involve co-polarized signal and idler photons, while in figure \ref{fig:PMContours}  (d) we show the phasematching contours for processes $3$ and $4$, involving cross-polarized signal and idler photons. 
Both figures suggest the potential for polarization entanglement two-photon state, especially in figure \ref{fig:PMContours} (d), around a $0.78\,\mu$m pump wavelength, for which the two differently-colored contours are nearly coincident; we have verified (not shown here) that scaling up the transverse fiber structure beyond the 10\% assumed here, tends to bring these two contours even closer together.
Note that panels (a)-(d) show that this fiber fulfills phasematching for all the processes summarized in table \ref{tab:process}, at pump wavelengths accessible from a Ti: Sapphire laser. 

Figure \ref{fig:PMContours} (e) shows the phasematching contour for the SFWM interaction in which the pump fields propagate in the $TM_{01}$ mode, while signal and idler fields travel in the $TE_{01}$ mode; this is similar to process 6 in table \ref{tab:process}, but involving the non-fundamental transverse modes.
We will leave discussion of panel (f), which involves counterpropagating SFWM, to section \ref{SFWMvariation}, subsection \ref{sec:counter}.

Figure \ref{fig:PMContours} and the discussion in this section makes it clear that optical fibers or waveguides, with a $\chi^{(3)}$ non-linearity, constitute a versatile platform for the generation of SFWM photon pairs characterized by a wide range of characteristics, as permitted by the dispersion properties. In this section, we have discussed how the spectral, transverse mode, and polarization degrees of freedom become intertwined through the phasematching properties, leading to the experimenter's ability to tap into any one degree of freedom, or any combination of degrees of freedom.  In particular, we will discuss in section \ref{sec:factorability} how the state can be engineered spectrally to be anywhere from factorable to highly entangled.

\section{The inner and outer branches of the phasematching diagrams}


In general, for SFWM processes involving all waves propagating in the same transverse and polarization modes, the solutions to $\Delta k=0$ in $\{\omega_p,\Delta_{s, i}\}$ space will be in the form of two symmetric loops (for the idler photon with $\Delta>0$ and for the signal with $\Delta<0$), provided that the pump spectral window of interest is large enough to encompass two zero dispersion wavelengths.   We can then distinguish between the portion of the loops furthest from the $\Delta=0$ axis (referred to as the outer branches) and the portions closest to the $\Delta=0$ axis (referred to as the inner branches).  In Fig. \ref{fig:inner_outer} we illustrate these two branches for the NL-PM-750 fiber with all four waves propagating in the $HE_{11}^x$ mode; while the solid black line represents the outer branch, the solid red line represents the inner branch.

For the former, signal and idler photons can typically be sufficiently separated from the pump field, to avoid correlation degradation of the two-photon states due to spontaneous Raman scattering.  It has been shown that the inner branches are particularly responsive to self/cross modulation effects mediated by the pump power (this effect is described in the next subsection), so that an increase in pump power leads to an enhanced separation between the inner branch and the $\Delta=0$ axis.  Thus,  it has been shown that through fiber dispersion management and pump power control it is possible to generate inner-branch photon pairs  sufficiently separated from the pump and the Raman background, even at low levels of pump power levels \cite{Garay2011a}.   Note that such Raman suppression is of interest because the inner branches lead to straightforward access to the group velocity matching regimes required for factorable and positive-correlation states (see further discussion of this in section \ref{sec:factorability}) \cite{Garay2011a}.



\section{Spectral shifting due to self/cross phase modulation}

In $\chi^{(3)}$ materials, nonlinear refraction effects, particularly self- and cross-phase modulation (SPM and XPM), add a term to the phase-mismatch which governs the SFWM process, which ultimately can lead to spectral shifting of the photon pairs. 
Note that since the pump waves are much more intense than the signal and idler fields, the latter have a negligible self phase modulation effect, or cross-modulation on other waves.
In consequence, the nonlinear term in the SFWM phase-mismatch relation is linear with the pump power and can be expressed as $\phi_{nl}=\gamma_1P_1+\gamma_2P_2$, where $\gamma_{1,2}$ is the fiber effective nonlinear coefficient and $P_{1,2}$ the pump peak power for each of the two pump waves, so that the phase mismatch becomes $\Delta k=k_{p1}+k_{p2}-k_{s}-k_{i}+\phi_{nl}$.

In figure \ref{fig:inner_outer} we show the effect of SPM and XPM on the SFWM phasematching properties of the NL-PM-750 fiber in a configuration in which all fields propagate in the $HE_{11}^x$ mode.  Dashed black lines indicate the effects of varying the pump power on the SFWM phasematching contours, with the blue arrows indicating an increasing pump power.
As can be seen, the phasematching contours shrink as the pump power increases, with the change particularly apparent in the position of the inner branches discussed in the previous subsection.   Note that if the pump power is further increased, eventually the phasematching loops shrink to a single point, and beyond that phasemathcing is no longer fulfilled (the pump powers typically used are significantly below this level).  Note that by increasing the nonlinear coefficients $\gamma_{1,2}$, a separation of the inner branches that is sufficient for Raman noise suppression occurs at lower pump powers. 

    \begin{figure*}[t!]
  \begin{center}
  		\includegraphics[width=0.8\linewidth]{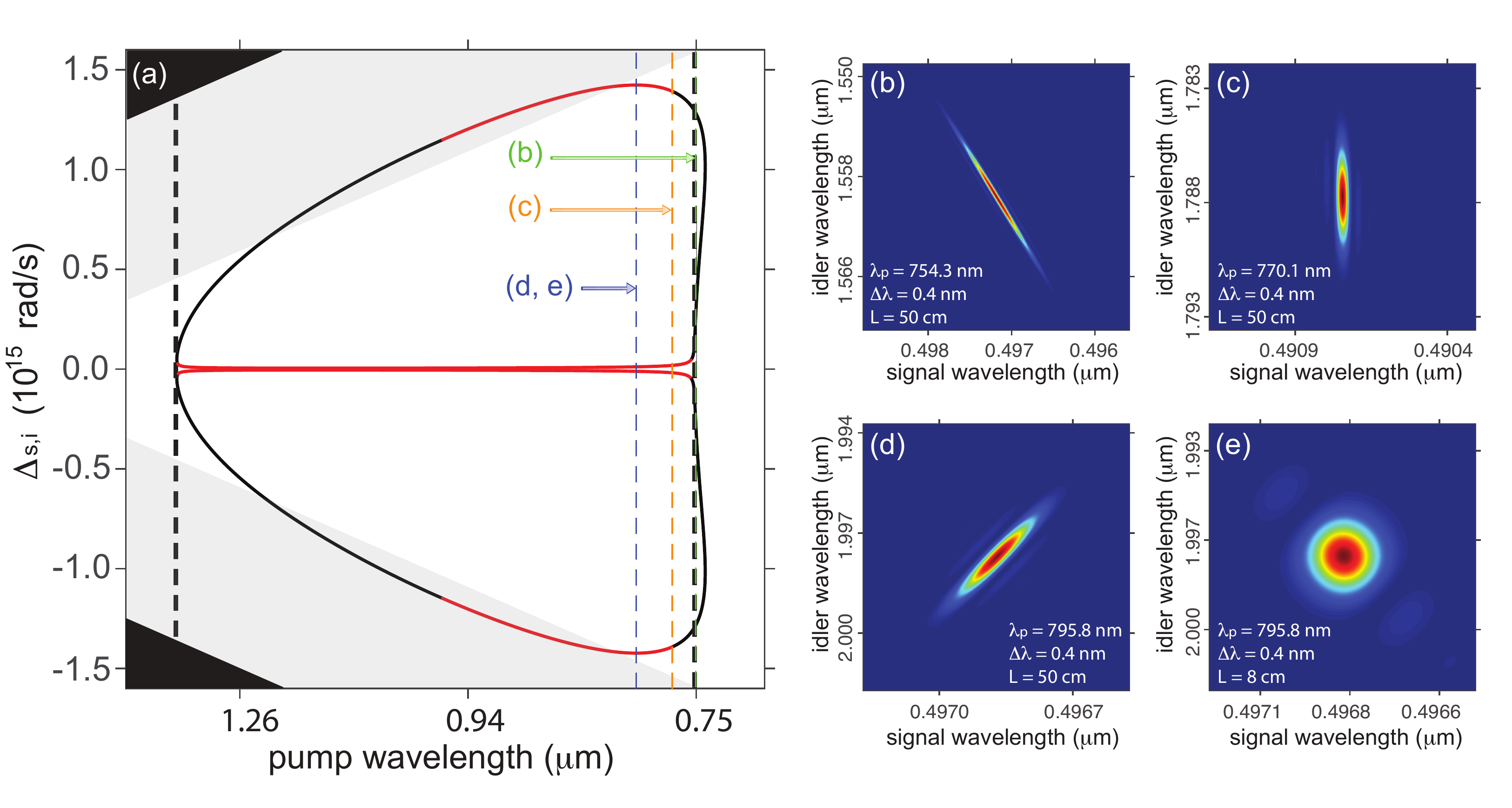}
  	\caption{(a) SFWM phasematching contour for the  NL-PM-750 fiber. Red portions of the diagram represent the wavelengths that fulfills, additionally to phasematching, group velocity matching properties for generating factorable two-photon states. (b)-(e) four different joint spectral intensities obtained for different choices of pump wavelength (indicated by vertical dashed lines in panel (a) along with a different choice of fiber length and pump bandwidth, as indicated in each panel).}
  	\label{fig:factregion}
  	\end{center}
  \end{figure*}

\section{Pure-state heralded single photons and factorability}
\label{sec:factorability}
Based on a SFWM photon-pair source, detection of one of the daughter photons (say in the idler mode) may herald the conjugate photon (in the signal mode) --- we refer to this arrangement as a heralded single-photon source.   It is well known that the purity of the heralded single photon $p_s$ is determined by the entanglement in the photon pair, as quantified by the Schmidt number $K$, as $p_s=1/K$ \cite{URen2005}.  This implies that in order to herald a quantum-mechanically pure single photon, the two-photon state produced by the SFWM source must be factorable.   Specifically for a single-mode fiber or waveguide, for which photon-pair entanglement may reside only in the spectral degree of freedom, this means that the joint spectrum must be factorable, i.e. functions $S(\omega_s)$ and $I(\omega_\textit{\textcolor{blue}{i}})$ must exist so that the joint spectrum  $F(\omega_s,\omega_i)$ may be written as $F(\omega_s,\omega_i)=S(\omega_s)I(\omega_i)$.

As described above in Eq.~\ref{eqn:jsalin}, the joint spectrum $F(\omega_s,\omega_i)$ may be expressed (within a small frequency interval around the central emission frequencies $\omega_{s0}$ and $\omega_{i0}$) as the product of the phasematching function $\vert \phi(\nu_s,\nu_i) \vert^2$ and the pump envelope function $\vert\alpha(\nu_s,\nu_i)\vert^2$; here $\nu_{\mu}=\omega_{\mu}-\omega_{\mu 0}$ represents the frequency detunings, for the signal ($\mu=s$) and idler ($\mu=i$) modes.    While the orientation of the $\vert \alpha(\nu_s,\nu_i)\vert^2$ function is always $-45^\circ$
in $\{\omega_s,\omega_i\}$ space, the orientation $\theta_{\omega_s \omega_i}$ of the $\vert\phi(\nu_s,\nu_i)\vert^2$ function depends on the dispersion properties of the fiber or waveguide.    It is known that as long as $\theta_{\omega_s \omega_i}$ lies within the range $0\le\theta_{\omega_s \omega_i}\le90^\circ$, a combination of fiber length and pump bandwidth exists yielding a factorable state \cite{Garay-Palmett:07}; note that this approach often requires the use of an ultrashort, e.g. ps- or fs-duration, pump.

Expanding $ k\left( \omega_{\mu }\right) $ in a first-order Taylor series about
frequencies for which perfect phasematching is attained $\omega _{p1,p2,s,i}^{0}$, the linearized phase mismatch $\Delta k_{lin}$ becomes

\begin{equation}  \label{eq: ldelk}
L \Delta k_{lin} = L\Delta k^{\left(0\right)}+T_s \nu_s +T_i \nu_i,
\end{equation}

\noindent where the zeroth order term $\Delta k^{(0)}$, must vanish to guarantee phase-matching at these center frequencies. The coefficients $T_{\mu }$ are given by $T_{\mu }=\tau _{\mu }+\tau _{p}\sigma _{1}^{2}/(\sigma _{1}^{2}+\sigma _{2}^{2})$,  where $\tau _{\mu }$ represent group-velocity mismatch terms
between the pump centered at frequency $\omega _{p2}^{0}$ and the generated photon centered at the frequency $\omega _{\mu}^{0}$, and $\tau _{p}$ is the group velocity mismatch between the two pumps

\begin{eqnarray}  \label{eq: taus}
\tau_{\mu} = L\left[k_{2}^{(1)}\left(\omega_{p2}^{0}\right) -
k_{\mu}^{(1)}\left(\omega_{\mu}^{0}\right)\right],  \notag \\
\tau_p = L\left[k_{1}^{(1)}\left(\omega_{p1}^{0}\right) -
k_{2}^{(1)}\left(\omega_{p2}^{0}\right)\right],
\end{eqnarray}

\noindent written in terms of $k_{\mu}^{(n)}\left(\omega
\right)=d^{n}k_{\mu}/d\omega^{n} |_{\omega=\omega_{\mu}^{0}}$ .

The type of spectral correlations observed in a SFWM two-photon state is determined in part by the slope of the phase-matching contour. If the phase-matching contour is given by a closed loop, all phase-matching orientation angles $\theta_{\omega_s \omega_i}$ are possible, controlled by the pump frequency. Thus, for certain relative orientations and widths of these two functions, it becomes possible to generate
factorable two-photon states. A factorable
state is possible if

\begin{equation}  \label{eq: CoDP_i}
T _{s}T _{i} \leq 0.
\end{equation}

Among those states which fulfil Eq.~\ref{eq: CoDP_i} those exhibiting a phasematching angle of $\theta_{\omega_s \omega_i}=45^{\circ}$, or $T_s=-T_i$, are of
particular interest.  For these states, in the degenerate pumps case, a factorable,
symmetric state is guaranteed if

\begin{equation}  \label{eq: CoDP_ii}
2\Gamma \sigma ^{2}|T _{s}T _{i}|=1,
\end{equation}

\noindent with $\Gamma \approx 0.193$. The condition in Eq.~\ref{eq: CoDP_i} constrains the group velocities: either $k^{(1)}(\omega _{s})<k^{(1) }(\omega _{p})<k^{(1) }(\omega_{i})$, or $k^{(1) }(\omega _{i})<k^{(1) }(\omega _{p})<k^{(1)}(\omega _{s})$ must be satisfied. The condition in Eq.~\ref{eq: CoDP_ii} constrains the bandwidths and fiber length. Thus, the region
in $\{\omega_s,\omega_i\}$ space in which factorability is possible is bounded by the conditions $T_{s}=0$ and $T _{i}=0$.

As has been described in Section \ref{PM}, a useful graphical depiction of the phasematching properties involves plotting the $\Delta k=0$ contour on the $\{\omega_p$,$\Delta_{s,i}\}$ space.   The slope of the $\Delta k=0$ contour $\theta_{\Delta \omega_p}$ is directly related to the joint spectrum angular orientation $\theta_{\omega_s \omega_i}=-\mbox{arctan}(T_s/T_i)$ through the relationship $\theta_{\Delta \omega_p}=45^\circ-\theta_{\omega_s \omega_i}$, which in turn means that within the range  $-45^\circ \le \theta_{\Delta \omega_p}  \le 45^\circ$ it becomes possible to obtain a factorable state. 

In figure \ref{fig:factregion}(a) we show a plot of the $\Delta k=0$ contour for a source based on the  NL-PM-750 fiber, assuming all waves propagate in the $HE_{11}^x$ mode, that exhibits two zero dispersion frequencies (indicated by vertical dashed lines in the figure). Note that the portion of the curve where factorable states are possible is highlighted in red.   In panels (b)-(e) we show four different examples of joint spectral intensities, each obtained for a different choice of pump wavelength, indicated with dashed vertical lines in panel (a), along with a different choice of fiber length and pump bandwidth.  Note that while panel (b) corresponds to a state with frequency anti-correlations and (d) to a state with positive correlations, both (c) and (e) represent factorable states which fulfill equation \ref{eq: CoDP_ii}.  It is noteworthy that the same fiber can produce a wide range of spectral correlation behavior. Note that the cases represented by the joint spectral intensities in panels d) and e) fulfill Eq.~\ref{eq: CoDP_i}, i.e. in both cases the  fields involved meet the group velocity matching condition; however, this only guarantees the orientation of the phasematching function. For the generation of a factorable two-photon state, Eq.~\ref{eq: CoDP_ii} must be satisfied additionally, which connects the pump bandwidth and fiber length, as is the case for the JSI in panel e).

\section{Photon-pair spectral tunability}\label{tunability}

As described above, particularly in the context of the $\Delta k=0$ contours in $\{\omega_p,\Delta_{s,i}\}$ space (see figures \ref{fig:PMContours} and \ref{fig:decres}), the SFWM phasematching properties depend on the underlying dispersion in the fiber or waveguide.   The overall dispersion is controlled, in turn,  on the one hand by the material dispersion, and on the other hand by the waveguiding geometry. The fact that as one decreases the core radius the waveguide dispersion component tends to become more dominant, can be used as the basis for spectrally tuning the two-photon state.  This idea has been experimentally demonstrated in Ref.~\cite{Ortiz_Ricardo_2017} in which it was directly shown in a SFWM photon-pair source based on an optical taper that the signal and idler emission frequencies can be controlled by the degree of tapering.  While the signal and idler were shown to be tunable over a range of $\sim 12-18$nm, for a limited tapered radius reduction (down  to $70\%$ of the original value), a much greater tuning range is expected for more drastic tapering (down to core diameters on the order of $1 \mu$m).   In this particular experiment, which relied on a bow-tie birefringent fiber, the maximum tapering was constrained by the onset of structural damage to the stress bows required to maintain polarization-maintaining behavior. As shown in the same paper, two possible alternatives to fiber tapering so as to achieve SFWM tunability (with a comparatively shorter tuning range), are fiber heating and the room-temperature application of longitudinal stress.  Note that a number of other works have demonstrated photon pair-sources based on micro/nano fibers or waveguides \cite{ 
fang2013,XiaoyingLi2018,Shin2019}.  As an example we show in figure \ref{fig:decres} the effect of decreasing the transverse dimensions of our NL-PM-750 fiber uniformly (i.e. a single scaling factor applied to the transverse structure). Panel (a) corresponds to our NL-PM-750 fiber, while panels (b) and (c) correspond to the phasematching contours for fibers with dimensions which are decreased by a factor of $10\%$ and $25.25\%$, respectively, with respect to fiber in (a). The figure 
clarifies that the phasematching contours both change in shape, as well as shift spectrally as a result of down-scaling the fiber transverse profile (note that the zero dispersion wavelengths shift from $0.75\mu$m and $1.39\mu$m to $0.71\mu$m and $0.91\mu$m).

\section{SFWM Bandwidth control}
In  SFWM two-photon states produced in single-mode fibers or waveguides, while entanglement in transverse modes is suppressed, one is relatively free to engineer the spectral degree of freedom.  In this case, the source can be designed to prepare single-photon wavepackets characterized by a spectral width ranging from ultra-broadband to ultra-narrowband. 

\subsection{Ultra-broadband photon pairs}

One of the key features of SFWM photon-pair sources is that through dispersion engineering of the fiber or waveguide it is possible to obtain a wide variety of two-photon states, ranging from ultra-narrowband to ultra-broadband, also including as has already been described factorable states.

The width of the signal-idler time of emission difference distribution is determined by the reciprocal of the SFWM emission bandwidth.   Therefore, an ultrabroadband photon-pair source leads to a high degree of simultaneity between the signal and idler photons (i.e. a narrow time of emission difference distribution).   Note that if the two emitted photons are input into a Hong-Ou-Mandel (HOM) inteferometer, such a sharp simultaneity is manifested as a very narrow HOM dip, which in turn serves as the basis for maximizing the attainable axial resolution in quantum optical coherence tomography (QOCT) setups \cite{Abouraddy2002}.  Note that fiber-based ultrabroadband photon-pair sources which are naturally collinear would be well-suited for Michelson-interferometer-based (rather than HOM-based) QOCT implementations \cite{LopezMago2012}.

\begin{figure*}[t!]
  \begin{center}
  		\includegraphics[width=0.9\linewidth]{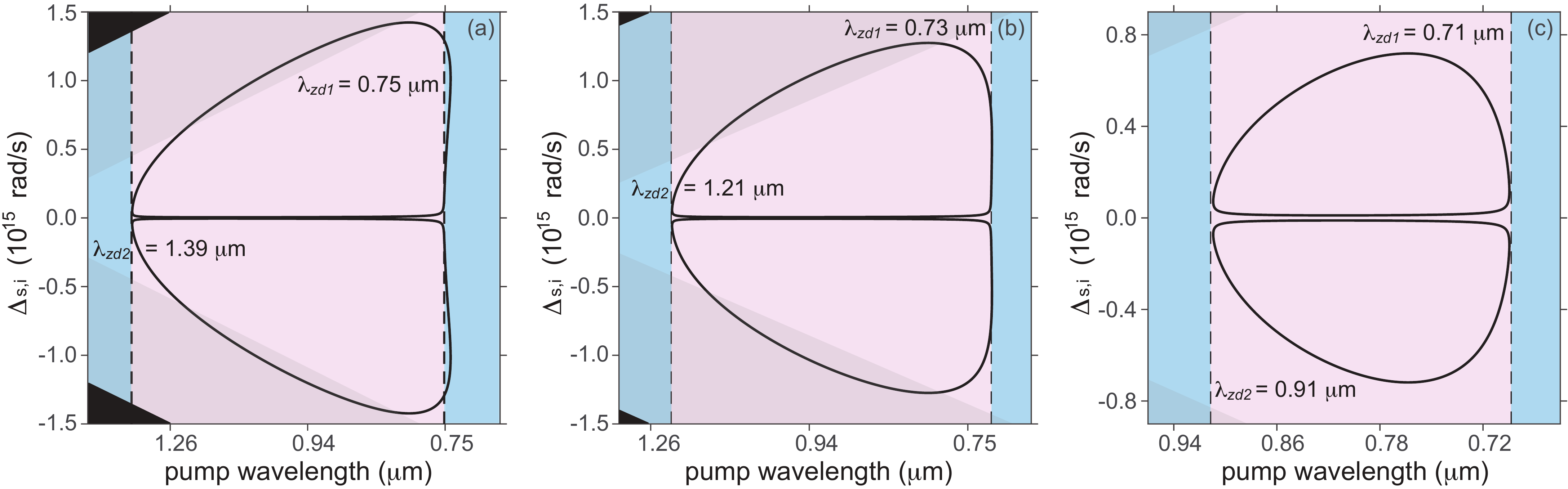}
  	          \caption{Phasematching diagrams for PCFs obtained by transverse scaling of the NL-PM-750 fiber. (a) Reference PCF. (b) Transverse dimensions decreased by $10\%$ with respect to the fiber in (a). (c) transverse  dimensions decreased by a $25.25\%$ with respect to fiber in (a). Blue(pink) areas represent normal (anomalous) dispersion regimes. All fibers exhibit two-zero dispersion wavelengths within the  spectral range of interest. The fiber in panel (b) fulfills the properties for ultra-broadband photon pairs generation.}
  	\label{fig:decres}
  	\end{center}
  \end{figure*}

In order to obtain ultra-broadband states, let us consider again the $\Delta k=0$ contours in $\{\omega_p,\Delta_{s,i}\}$ space (see figure \ref{fig:decres}).   As has already been discussed, along the $\omega_p$ axis, the phasematching curve is approximately bounded by the zero dispersion frequency or frequencies.    What is needed for the state to be ultrabroadband is that for a fixed value of $\omega_p$ the longest possible stretch of $\pm\Delta_{s,i}$ values (representing the signal and idler frequencies) are phasematched.  This can be accomplished  by choosing a fiber geometry which results in the suppression  of  the $\Delta k=0$ contour curvature at $\Delta_{s,i}=0$.  As discussed in Ref. \cite{Garay2008}, this curvature is proportional to $k^{(4)}/k^{(3)}$, where $k^{(n)}$ represents the $n$th frequency derivative of $k$.  The condition $k^{(4)}=0$ at the zero dispersion frequency therefore turns into the condition $k^{(2)}=k^{(4)}=0$, i.e. both the second and fourth derivatives must vanish.

As explained in section \ref{tunability}, figure \ref{fig:decres} illustrates the explained behavior. Panel (a) corresponds to the  NL-PM-750 fiber, while panels (b) and (c) show the phasematching contours for fibers with dimensions decreased by a factor of $10\%$ and $25.25\%$, respectively, with respect to the  NL-PM-750 fiber. It can be seen that the fiber considered for panel (b) meets the properties for ultra-broadband photon pair generation, as discussed in reference \cite{Garay2008}.

\subsection{Ultra-narrowband photon pairs}

On the opposite extreme, SFWM  sources may be designed and built so that the emitted photon pairs are characterized by an ultra-narrow bandwidth.   

A heralded single photon, obtained from a SFWM photon-pair source, may serve as a flying qubit for quantum information processing applications.    Provided that such a single photon can be deterministically absorbed by a single atom, the latter may then constitute a quantum memory.  Such efficient single-photon absorption by a single atom requires matching: i) the single photon frequency to the desired electronic transition of the atom, and ii) the single photon bandwidth to that of the transition.    

Note that for spontaneous parametric processes, including both SPDC and SFWM,  there is no intermediate level defining the energy splitting ratio for the signal and idler photons.  The range of possible splitting ratios, which determines the signal and idler emission bandwidth, is mainly limited by the phasematching constraint, which can be many orders of magnitude larger than the typical electronic transitions with bandwidths on the order of MHz.  In order to design SFWM sources which can emit photon pairs with MHz or sub-MHz bandwidths two main strategies are possible:  i) the use of atomic non-linearities as the basis for SFWM which are naturally narrowband and directly suited for single photon - single atom interfaces \cite{Shu2016,Park2017}, ii) the use of cavity-enhanced SFWM in which photon pair emission occurs only within the sharp spectral resonances defined by resonator \cite{Garay2012}.  The first strategy is promising but lies outside the scope of the fiber-based focus of this tutorial paper. The second strategy includes a considerable body of work based on nonlinear devices in the form of a micro-cavities including disks \cite{Kippenberg2006}, toroids \cite{Spillane2005}, and spheres \cite{OrtizRicardo2021}, all with large $Q$ values, which likewise lies outside the current fiber-based focus. An interesting possible approach is the use of extended fiber cavities, either of a linear kind based on fiber Bragg resonators to act as mirrors, or of a loop kind with one output port of a fiber beamsplitter connected to one input port, and the two remaining ports acting as overall input and output for the device. Such a source could provide an interesting platform for the generation of narrowband photon pairs for single photon -- single atom interfaces.

\section{SFWM transverse mode control}

As explained above, the SFWM process photon-pair emission can include a coherent superposition of a number of different processes, each characterized by a particular combination of transverse modes and polarizations for the four participating fields (see Eq. \ref{eqn:state} and figure \ref{fig:PMContours}). Thus, the two-photon state can include in this manner contributions from various transverse modes.   With the aim of taking advantage of the transverse-mode degree of freedom, it is necessary to: i) control the transverse modes present in the emitted state, ii) be able to deterministically switch from one emitted mode to another, and iii) be able to transmit over longer distances single photons described by particular transverse modes. 

In Ref. \cite{LeonSaval2014} it was demonstrated that a device known as a mode selective photonic lantern (MSPL) can accomplish the deterministic conversion of the transverse mode in which a single photon travels to the coherent superposition of two other modes.  Note that the polarization degree of freedom has been used widely in order to encode information, in part because of the readily available optical elements such as polarizers and waveplates which permit the full manipulation of this degree of freedom.  It is comparatively more difficult to control the transverse mode degree of freedom, due to the unavailability of devices which act in an analogous manner to polarizers and waveplates. In Ref. \cite{Cruz-Delgado2019}  it is shown that a MSPL device may act as a half-wave plate analogue for spatial modes, and that the mode-converted single photon may travel over a long distance in a few-mode fiber with minimal mode distortion.   

\section{Spontaneous Raman Scattering as a noise mechanism}
\label{sec:raman}

When compared to photon-pair sources based on spontaneous parametric downconversion (SPDC), those based on SFWM in fibers tend to be affected to a greater degree by noise mechanisms.   The appearance of uncorrelated noise photons has the effect of weakening the signal-idler correlations, which can be quantified through a reduced coincidence-to-accidental ratio (CAR). Among possible noise mechanisms affecting SFWM photon pair sources, spontaneous Raman scattering (SRS) of the pump field or fields plays an important role \cite{Fan_J2007}.  Particularly for fused-silica optical fibers, the dominant Stokes contribution to  SRS occurs towards the red from each of the pumps over a bandwidth of $\sim 50$ THz. Photon-pair sources based on other materials such as chalcogenides \cite{Clark2012,Collins2012}  have also been shown to be affected by SRS.  Possible strategies to mitigate of the SRS contamination of emitted photon pairs include: i) relying on a phasematching scheme with a large spectral detuning between the pump and each of the signal and idler photons (particularly relying on the outer branch of the phasematching curves; see figure \ref{fig:inner_outer}), ii) cryogenic cooling of the optical fiber \cite{Takesue2005}, and iii) using SFWM sources in which the nonlinear material is in gaseous form, e.g. in xenon-filled hollow-core fibers \cite{Larson2022}, or in liquid form as in references \cite{Barbier_2015,Afsharnia:22}.

\section{SFWM process variations} \label{SFWMvariation}

The two-photon state given by equation \ref{eqn:state} and the phasematching properties examined in section \ref{PM} highlight the variability of configurations in which the SFWM may be implemented, leading to the generation of photon pairs with engineered entanglement properties involving discrete (polarization, spatial mode) and continuous (spectral) degrees of freedom. This section discusses SFWM variations implemented in different optical fibers and beyond the co-polarized, co-propagating, and single-mode configuration.

\subsection{SFWM in various types of fiber} \label{sec:fiberTypes}
Since the first demonstration of an SFWM source \cite{Fiorentino2002}, several fiber geometries have been used to exploit this nonlinear interaction's potential for preparing two-photon states with engineered correlation and entanglement properties, as discussed in previous sections. Fortunately, fiber optic technology has advanced dramatically in the last two decades \cite{Addanki2018,Pisco2020,Yonggang2020,Li:21}, leading to the accessibility of a system with controllable dispersion properties that can be exploited in quantum optics to engineer photon-pair sources for different applications. 
To mention a few, the types of optical fibers that have been used for the implementation of photon pairs are telecom dispersion-shifted \cite{Li:04,Takesue:09,Liu:16}, polarization-maintaining \cite{Smith:09,Zhou2014,Cruz-Delgado2016}, commercial grade \cite{Soller2011}, photonic crystal \cite{Rarity:05,Fulconis2007,McGuinness2010,Cui_2012,Daniel2021}, hollow-core \cite{Barbier_2015,Cordier2020,Larson2022}, tapered \cite{Meyer-Scott2015,Ortiz_Ricardo_2017,Shukhin2020}, and few-mode fibers \cite{Garay16,Rottwitt2018,Guo:19,Shamsshooli2021}.

\subsection{Non-degenerate dual pump scheme} \label{sec:dualpump}
 
As opposed to spontaneous down-conversion (SPDC), which relies on a $\chi^{(2)}$ nonlinear interaction in which only one pump photon is annihilated, in SFWM, the $\chi^{(3)}$ nonlinear process involves the annihilation of \textit{two} pump photons and the creation of one signal and one idler photon. The dual pump nature of SFWM makes the photon-pair generation process more controllable through different pump polarizations \cite{Li2005,Fan_J2007,Kaiser_2012,Fang2014}, transverse modes \cite{Cruz-Delgado2014,Nazemosadat2016,Cruz-Delgado2016}, frequencies \cite{Pourbeyram2015,Daniel2021}, and/or temporal delay \cite{fang2013,Zhang2019}, which can lead to a variety of applications including creating factorable, ultrabroadband, or ultranarrowband photon-pair states, as discussed in previous sections.

Without special treatment, the photon pairs produced in SFWM are in general entangled in the spectral and spatial degrees of freedom due to the strong correlations introduced by energy
and momentum conservation constraints. To reduce the spectral entanglement in the photon-pair generation, much progress has been made in tailoring the photon-pair joint spectrum by using degenerate pumps and shaping pumps temporally (see section 8) \cite{Garay-Palmett:07}. In this section, we discuss how SFWM with two frequency-non-degenerate pumps with temporal walk-off in standard silica polarization-maintaining fibers (PMFs) can be used to eliminate the spectral correlation.

In the weak pump power regime where self- and cross-phase modulation can be ignored, the SFWM process is constrained by the phasematching conditions \cite{Garay-Palmett:07,fang2013}
\begin{align}
&\Delta\omega=\omega_{p1}+\omega_{p2}-\omega_s-\omega_i=0\\
&\Delta k=k_{p1}(\omega_{p1})+k_{p2}(\omega_{p2})-k_{s}(\omega_{s})-k_{i}(\omega_{i})=0.
\end{align}
When considering the case where the pumps' polarizations are along one of the principal axes of the fiber, while the signal and idler photons are generated orthogonally to the pump, the dispersion relation will be given as $k_{p1}(\omega)=k_{p2}(\omega)=\frac{\omega(n(\omega)+\Delta n)}{c}$ and $k_{s}(\omega)=k_{i}(\omega)=\frac{n(\omega) \omega}{c}$, where $\Delta n$ is the fiber birefrengence.
Expanding on the first-order linear approximation for the joint spectral function introduced in section 1 to include the pump spectral envelopes, we have 

\begin{equation}
\begin{split}
  F_{lin}(\nu_s,\nu_i)&=\alpha(\nu_s+\nu_i)\phi(\nu_s,\nu_i)\\
&=\exp{\left[-\frac{(\nu_s+\nu_i)^2}{\sigma_1^2+\sigma_2^2}\right]}\exp{\left[-\left(\frac{T_s\nu_s+T_i\nu_i}{\sigma\tau_p}\right)^2\right]}\\
&\times\Bigg[\erf\left(\frac{\sigma(\tau+\tau_p)}{2}-i\frac{T_s\nu_s+T_i\nu_i}{\sigma\tau_p}\right)\\
&\quad-\erf\left(\frac{\sigma\tau}{2}-i\frac{ T_s\nu_s+T_i\nu_i}{\sigma\tau_p}\right)\Bigg],
\end{split}
\end{equation}
where $\sigma_{1(2)}$ denotes the pump $1(2)$ spectral bandwidth while $\sigma=\sigma_1\sigma_2/\sqrt{\sigma_1^2+\sigma_2^2}$ is the effective pump bandwidth;  $T_{\mu}=\tau_{\mu}+(\sigma_1^2-\sigma_2^2)/(\sigma_1^2+\sigma_2^2)\tau_p/2$, where  $\tau_{\mu}=L((k'_{p1}+k'_{p2})/2-k'_{\mu})$ is the group delay difference between the signal (idler) and the average group delay of the pumps acquired during the propagation in the fiber, where $k'_{\mu}=dk/d\omega|_{\omega_{\mu}}$ is the inverse group velocity of the signal/idler in the fiber, $\tau_p=L(k'_{p1}-k'_{p2})$ is the group delay between the two pumps acquired during the propagation in the fiber, and $k'_{p_{1(2)}}=dk/d\omega|_{\omega_{p_{1(2)}}} + \Delta n/c$ is the inverse group velocity of pump $1(2)$ in the fiber.

As mentioned in section~\ref{sec:factorability}, since the pump envelope function $\alpha(\nu_s+\nu_i)=\exp{\left[-\frac{(\nu_s+\nu_i)^2}{\sigma_1^2+\sigma_2^2}\right]}$ has an orientation of $-45^{\circ}$ in $\{\omega_s,\omega_i\}$ space, in order to obtain a factorable state, the phase-matching function angle $\theta_{\omega_s\omega_i}=-\arctan(T_s/T_i)$ should be set to be in the range $0^{\circ}< \theta_{\omega_s\omega_i}< 90^{\circ}$, which corresponds to the condition $T_sT_i<0$. There are two regions in which this condition can be satisfied

\textit{Negligible temporal walk-off}: When the temporal walk-off between the pumps is negligible (i.e. $|\sigma\tau_p|\ll 1$ or even $\tau_p=0$). This is the case discussed in section~\ref{sec:factorability}, and in Ref.~\cite{Garay-Palmett:07}, where the phase-matching function is reduced to a sinc function $\phi_{|\sigma\tau_p|\ll 1}(\nu_s,\nu_i)=\text{sinc}(\frac{T_s\nu_s+T_i\nu_i}{2})$ with $T_s\approx -T_i$. In this case, the phase-matching function has an orthogonal orientation compared to the pump envelope function $\alpha(\nu_s+\nu_i)$; however, unlike the Gaussian function, the
oscillatory behavior of the sinc function carries sidelobes that result in spectral correlation and thus limits the maximal factorability that one can achieve.

\begin{figure*}[t]
 \center
  \includegraphics[width=\textwidth]{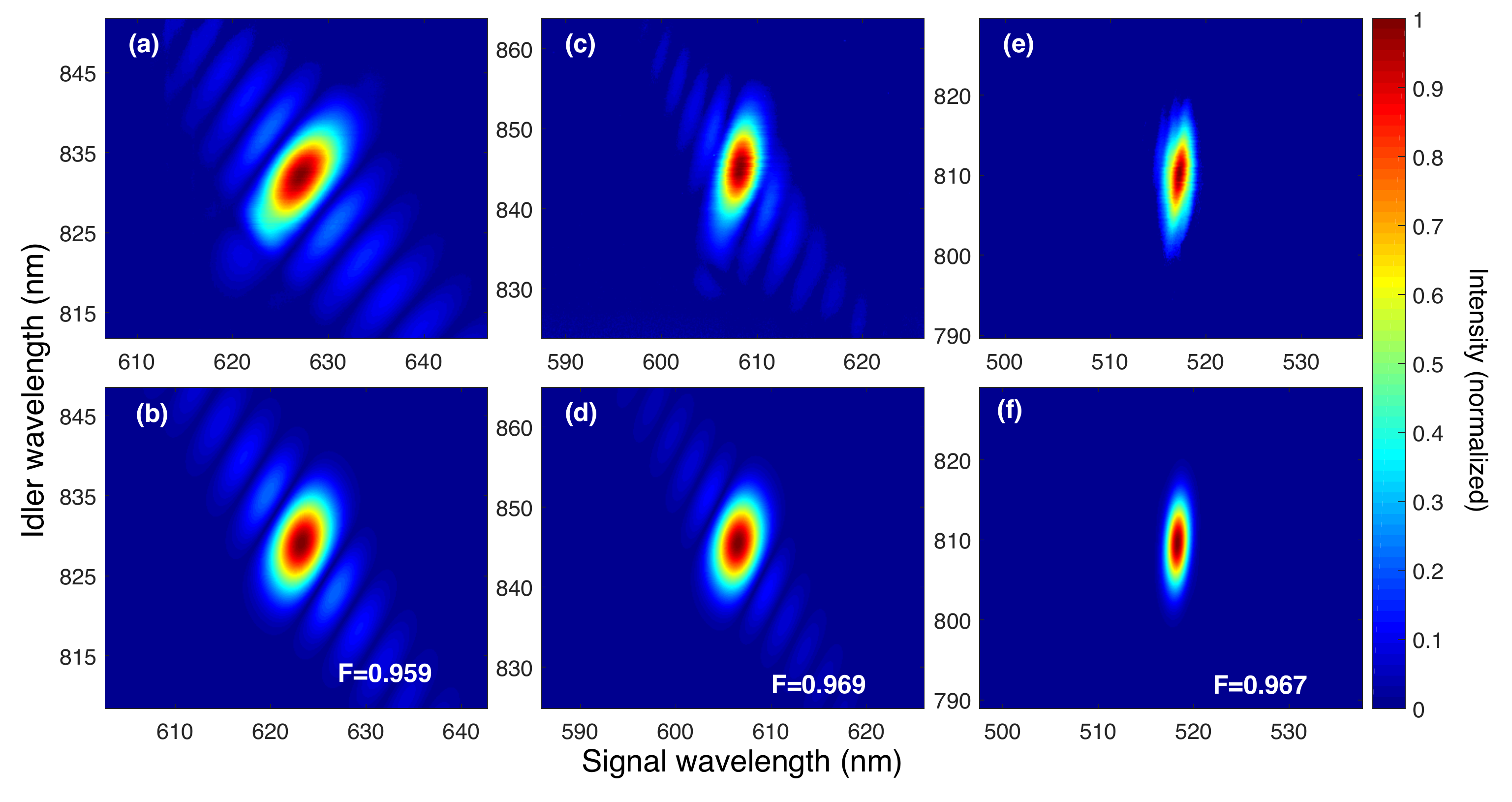}
  \caption{Experimental (top row) and theoretical (bottom row) joint spectral densities (JSDs) for various detunings. Experimental data is measured via stimulated emission. (a,b) degenerate pump at $715\,$nm, (c,d) dual pump at $772\,$nm and $652\,$nm, (e,f) dual pump at $772\,$nm and $534\,$nm. Going from left to right, corresponding to increasing detuning between the two pumps, the sidelobes' intensity weakens and the JSD of the signal and idler photons becomes less correlated. Here, $F$ is the fidelity of the experimental JSD with respect to the corresponding theoretical JSD~\cite{Zhang2019}.}
  \label{fig:jsi}
\end{figure*}

\textit{ Complete temporal walk-off}: The opposite limit, where the two pumps are temporally separated  with $|\sigma\tau_p|\gg 1$. In this region, when sending the slow pump ahead
of the fast by a time $|\tau_p/2|$, the interaction gradually increases as the fast pump catches up with the slow pump and turns off and vanishes as the pumps separate towards the end of the fiber, which results in an interaction function of a Gaussian shape (depending on the pump envelope function) and thus the phase-matching function can be approximated to be a Gaussian function
\begin{equation}
    \phi_{|\sigma\tau_p|\gg 1}(\nu_s,\nu_i)=\exp\left[-\left(\frac{T_s\nu_s+T_i\nu_i}{\sigma\tau_p}\right)^2\right].
\end{equation}
In PMF, when working at regions far from the fiber zero dispersion wavelength (i.e $\tau_s=-\tau_i$),  factorability is possible when the pumps are well-separated (i.e $\tau_p\approx \frac{\tau_s}{2}$). A numerical simulation as well as experimental demonstration can be found in Fig.~\ref{fig:jsi} \cite{Zhang2019}.  Note that the choice of non-degenerate dual pump configuration determines the levels of factorability that can be achieved.

\subsubsection{Counter-propagating SFWM photon pair sources} \label{sec:CPpumps}
\label{sec:counter}

In all SFWM sources demonstrated to date, all four participating optical fields, i.e. the two pumps, signal, and idler, propagate along the nonlinear medium in the same direction. Specifically in relation to fiber- and waveguide-based photon-pair generation, an interesting possibility is to design sources in which the two pump fields (with frequencies $\omega_1$ and $\omega_2$) are launched into the nonlinear medium from opposite ends so that they counter-propagate \cite{Saravi2022}.  It has been shown \cite{Monroy_Ruz_2016} that, in a situation in which all four fields are co-polarized and propagate in a single transverse mode, phasematching is then automatically attained in such a manner that the signal photon is emitted with frequency $\omega_s=\omega_1$ in a direction opposite to that of pump 1, and the idler photon is emitted with frequency $\omega_i=\omega_2$ in a direction opposite to that of pump 2.  Such a process is referred to as counter-propagating SFWM (CP-SFWM).  In addition, in Ref. \cite{Monroy_Ruz_2016} it has been shown that if the region of the fiber (or waveguide) in  which the two pumps overlap ---assuming that both pumps are pulsed--- is much shorter than the fiber length, the state becomes \emph{automatically factorable} (regardless of the material dispersion properties), thus ensuring that single heralded photons derived from such a source are in a quantum-mechanically pure state. For the mentioned pump scheme, the joint spectral amplitude ($F_{CP}$) of the two-photon state is given by

\begin{equation}
	\begin{aligned}
      F_{CP}(\omega_s,\omega_i)&=\int\! d\omega\,\alpha_{1+}(\omega)\alpha_{2-}(\omega_s+\omega_i-\omega)\mbox{sinc}\left[\frac{L}{2}\Delta k_{CP}\right]\\ \nonumber
 &\times\mbox{exp}\left[i\frac{L}{2}\kappa\right]\mbox{exp}(i\omega\tau),
	\end{aligned}
	\label{eqn:jsaCounter}
\end{equation} 

\noindent where $\alpha_{1+}(\omega)$ represents the pump spectral envolope for the forward-propagation pump, while $\alpha_{2-}(\omega)$ is the pump spectral envelope for the backward-propagating pump. $\tau$ represents the time arrival difference between the two pump pulses at the corresponding fiber end. In this case, the phase mismatch function $\Delta k_{CP}\equiv\Delta k_{CP}(\omega,\omega_s,\omega_i)$, and the function $\kappa\equiv\kappa(\omega,\omega_s,\omega_i)$ are given by

\begin{equation}
	\begin{aligned}
      \Delta k_{CP}=k_1(\omega) - k_2(\omega_s+\omega_i-\omega) - k_s(\omega_s)+k_i(\omega_i)+\phi_{NL} ,
	\end{aligned}
	\label{eqn:DKcp}
\end{equation} 

\begin{equation}
	\begin{aligned}
      \kappa=k_1(\omega) + k_2(\omega_s+\omega_i-\omega) + k_s(\omega_s)+k_i(\omega_i),
	\end{aligned}
	\label{eqn:kappa}
\end{equation} 

\noindent with $\phi_{NL}$ the phase shift related to processes derived from the nonlinear refraction effect \cite{Monroy_Ruz_2016}.

The key importance of this proposal is that while in general for a given nonlinear optical material phasematching and factorability (group velocity matching) occur only for specific sets of wavelengths, CP-SFWM permits these conditions at arbitrary, user-controlled frequencies.   Note that in practice the $\omega_s=\omega_1$ and $\omega_i=\omega_2$ symmetry may be broken (aiding experimental pump-SFWM discrimination), by letting the four participating modes propagate in different spatial modes or with different polarizations, as shown in figure \ref{fig:PMContours} (f).

In figure \ref{fig:PMContours} (f) we show the phasematching characteristics for a counterpropagating  configuration, for the same fiber assumed in section \ref{PM}. In this case, we fix the pump-$2$ wavelength  at $0.532\,\mu$m while allowing the  pump-$1$ wavelength to vary. We have plotted in a yellow dashed line the phasematching  contour for the situation in which all  four fields  propagate in the $HE_{11}^x$ mode.  As can be seen, the $\Delta k=0$ contour is formed by two oblique lines that cross at the degeneracy point ($\omega_{1}=\omega_{2}=\omega_s=\omega_i=2\pi c/\mbox{0.532}\,\mu$m), with signal and idler photons traveling in counter-propagation. Other solutions within the yellow lines represent signal and idler pairs that counter-propagate with $\omega_1=\omega_s$ and $\omega_2=\omega_i$ (symmetric phasematching). Such symmetry can be broken if the SFWM process is allowed to involve  different polarizations and transverse modes, as shown in figure \ref{fig:PMContours}(f), where we have shown the $\Delta k=0$ contours for the following SFWM interactions: i) $+HE_{11}^y$ - $TM_{01}$  $\rightarrow$ $-TE_{01}$+$HE_{11}^x$, ii) $+HE_{11}^y$ - $TM_{01}$  $\rightarrow$ $+TE_{01}$-$HE_{11}^x$, iii) $+HE_{11}^y$ - $TM_{01}$  $\rightarrow$ $-HE_{11}^x$ + $TE_{01}$ and iv) $+HE_{11}^y$ - $TM_{01}$  $\rightarrow$ $+HE_{11}^x$ + $-TE_{01}$, with $+$($-$) representing forward(backward) direction. Note that the colors follow the convention used in section \ref{PM}.   It is remarkable in this figure that there is a phasematched solution that departs from  symmetric phasematching, in which the SFWM photons propagate in the modes $HE_{11}^x$  and $TE_{01}$,
plotted in  solid black and blue lines. Note that in this configuration, the counterpropagating pairs pump 1-signal and pump 2-idler become orthogonally polarized and acquire a spectral shift, aiding experimental discrimination.

\subsection{Quantum correlations in different degrees of freedom}
From equation \ref{eqn:state} and the discussion in section \ref{PM}, we know that the flexibility of the SFWM process can be exploited so as to generate 
photon pairs exhibiting correlations, and entanglement,  in different 
degrees of freedom, or combinations of degrees of freedom.



\subsubsection{Entanglement generation}

While correlations in continuous-variable degrees of freedom are sometimes undesirable, the intentional creation of correlations in energy-time \cite{Harada08,Dong2014}, polarization \cite{Li2005,Zhou:12,Fang2014,Karmakar:15}, spatial mode \cite{Cruz-Delgado2016,Ekici2020,Shamsshooli2021}, or discrete frequency \cite{Li09,Dong2015} is important for many applications in quantum information science. Fiber-based entangled photon pair sources hold the advantage of ready integration into fiber-based quantum networks and of supporting high-dimensional discrete transverse modes. We now consider fiber-based sources of photon pairs entangled in various degrees of freedom.

\textit{Entanglement in energy and time:} 
Although continuous spectral correlation in the continuous-wave pump regime is often undesirable for applications that require indistinguishable photons, such correlations can be a resource for other quantum applications. Photon pairs with continuous energy-time correlations can be understood as entangled in the infinite-dimensional Hilbert space and thus can violate the Clauser-Horne-Shimony-Holt (CHSH) inequality. Energy-time entanglement can be measured in a Franson-type interferometer consisting of two unbalanced Mach-Zehnder interferometers \cite{Dong2014,Dong2015,Fang2019}. Since the photon pair is strongly correlated in time, the resulting state can be written as:
\begin{equation}
    \ket{\Psi}=\frac{1}{\sqrt{2}}[\ket{\text{short}}\ket{\text{short}}+e^{i(\alpha+\beta)}\ket{\text{long}}\ket{\text{long}}],
\end{equation}
where $\alpha$ and $\beta$ are the phases that are applied in the unbalanced Mach-Zehnder interferometers. The unbalanced interferometer path length difference is set to be much shorter than the coherence time of the pump but longer than the coherence time of the signal and idler.

\textit{Entanglement in polarization:}
Polarization-entangled photon pairs can be created through SFWM inside a polarization-maintaining fiber (PMF) Sagnac interferometer \cite{Fang2014}. The pump, whose polarization is oriented at 45$^\circ$, is split equally by a polarizing beamsplitter. The horizontal component is transmitted and then coupled into one end of the PMF to generate photon pairs with vertical polarization, which changes to horizontal polarization due to a $90^{\circ}$ twist in the fiber. Similarly, the vertical pump component is coupled to the other end of the PMF and generates photon pairs with horizontal polarization, which exit the fiber vertically polarized. At the output port of the  beamsplitter the photon pair state generated is of the form:
   \begin{equation}
         \ket{\Psi}=\frac{1}{\sqrt{2}}\ket{\Phi}(\ket{HH}+e^{i\phi}\ket{VV}),
   \end{equation}
where $\ket{\Phi}$ is the spectral-temporal photon-pair state assuming pumps with fixed polarization along the principal axes of the fiber.

\textit{Entanglement in frequency-transverse-mode:}
If instead of working with single mode fiber as in the previous examples, few-mode or multi-mode fiber is employed, more than one spatial mode should be considered in the SFWM process, as is reflected in the general photon-pair state represented by Eq.~\ref{eqn:state}.
Different spatial modes in the fiber experience different refractive indices; therefore, generally speaking, photon pairs in different spatial modes are generated at different frequencies. This results in spatial-frequency correlations between the photons. The discrete nature of these correlations allows for their full characterization and can be used to build hyper-entangled or hybrid-entangled quantum states of the form $\ket{\psi}=\sum_n\ket{\omega^n_{s}}\ket{\omega^n_{i}}\ket{q^n_{s}}\ket{q^n_{i}}$, where $\ket{q^n}$ stands for different spatial modes \cite{Ekici2020}.

Few-mode optical fibers permit the generation of photon pairs with a quantum state given by the coherent superposition of distinct SFWM processes, each related to a different combination of transverse modes amongst the four waves involved. Such a source involving three different processes, based on a bow-tie birefringent fiber and on process 5 in table \ref{tab:process}, i.e. $xx-yy$,  was experimentally demonstrated in Ref. \cite{Cruz-Delgado2016}.  Note that by using a shorter pump wavelength, or larger core radius so that a greater number of transverse modes are supported, it is possible to obtain a state built as the coherent superposition of a larger number of distinct processes.  Such a source has the potential of exploiting hybrid entanglement in frequency and transverse mode.

\begin{figure}[t!]
  \begin{center}
  		\includegraphics[width=0.85\linewidth]{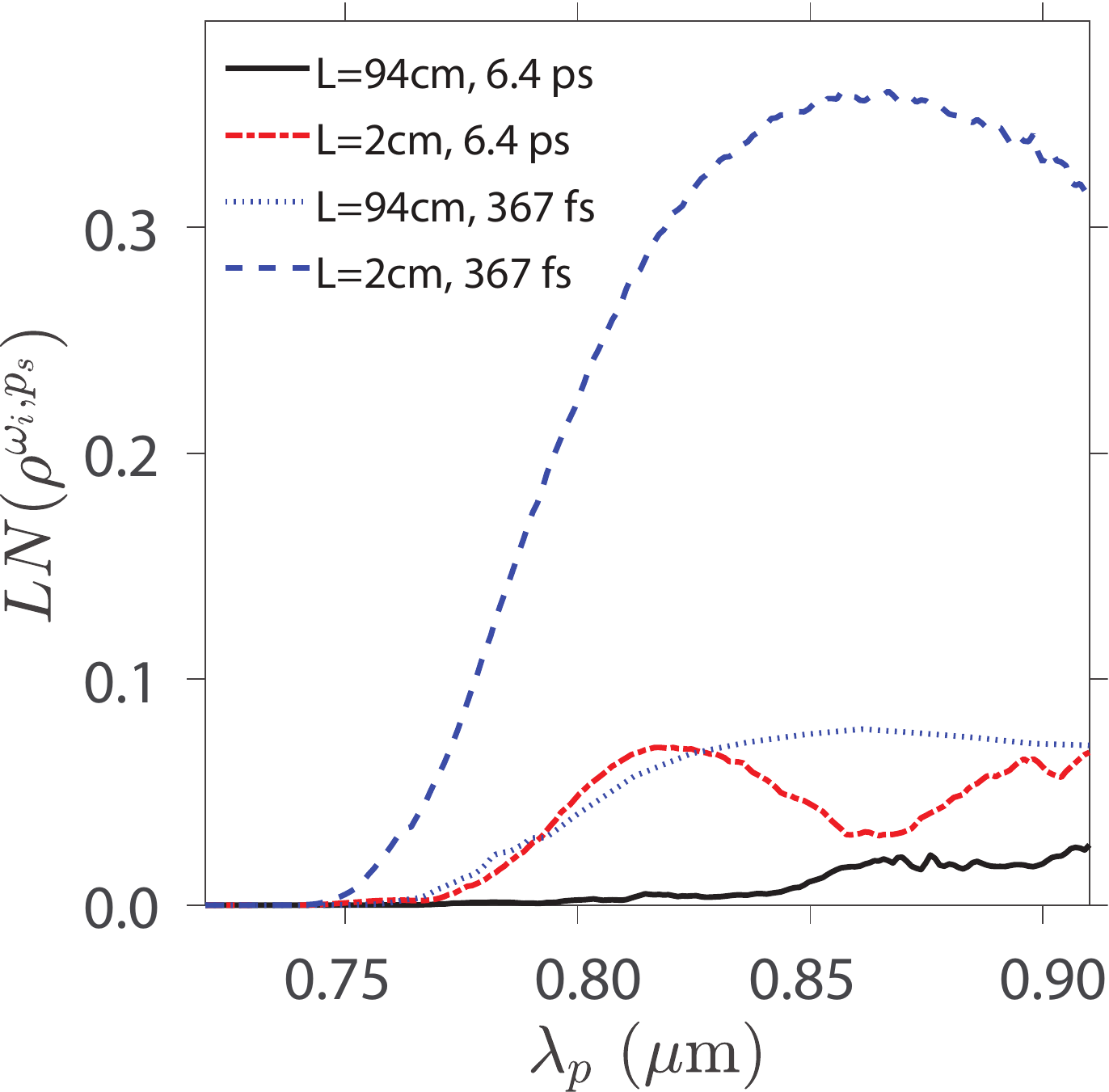}
	\caption{Logarithmic negativity as a function of the pump wavelength for different fiber lengths and pump pulse durations in the picosecond (ps) and femtosecond (fs) regimes \cite{Daniel2021}.}
  	\label{fig:Neg}
  	\end{center}
  \end{figure}

\textit{Entanglement in frequency-polarization:}
Single-transverse spatial mode fibers that support the two fundamental polarization modes permit the generation of two-photon states which result from the coherent superposition of SFWM interactions related to the different polarization combinations amongst the four participating optical fields.  In general, this leads to two-photon states with hybrid entanglement in the spectral and polarization degrees of freedom. The generation of photon pairs exhibiting hybrid frequency-polarization correlations in a photonic crystal fiber has been experimentally demonstrated in Ref. \cite{Daniel2021}. For this two-photon state we have calculated, in four different source configurations, the \textit{logarithmic negativity} (LN) as a metric of the hybrid frequency-polarization entanglement \cite{Vidal2002,Plenio2005}, particularly in the frequencies of the idler mode ($\omega_i$) and the polarization of the signal mode ($p_s$). Results are shown in figure \ref{fig:Neg}, where it can be seen that the  hybrid frequency-polarization entanglement increases for shorter fibers and shorter pulse duration, conditions that favor the spectral overlap among the processes corresponding to different polarization combinations, see table \ref{tab:process}.

\textit{Entanglement in discrete frequency mode:} Beyond bipartite entangled state generation via SFWM, the higher-order terms in SFWM can be used to create multipartite entangled states. An example is a recent experimental demonstration of tripartite-entangled W state generation in the discrete energy degree of freedom in optical fiber \cite{Fang2019}. In this case, the pump power is increased until the second-order term in SFWM becomes relevant
\begin{equation}
    \ket{\Psi}=(1-O(\beta^2))\ket{\text{vac}}+\beta a_{\omega_s}^{\dagger}a_{\omega_i}^{\dagger}\ket{\text{vac}}+\frac{\beta^2}{2} (a_{\omega_s}^{\dagger})^2(a_{\omega_i}^{\dagger})^2\ket{\text{vac}}.
\end{equation}
By sending the pairs to a series of three beamsplitters that are balanced for equal output probability, a three-photon W state on post-selection can be formed
\begin{equation}
    \ket{W}=\frac{1}{\sqrt{3}}(\ket{\omega_i\omega_s\omega_s}+\ket{\omega_s\omega_i\omega_s}+\ket{\omega_s\omega_s\omega_i}).
\end{equation}

\section{SFWM two-photon state characterization} \label{sec:characterization}

As discussed in the previous sections, optical fiber sources can be cleverly engineered to create photon pairs correlated in various degrees of freedom (DOF), and sometimes in complex combinations of them. Depending on the possible applications, we might want to exploit those photon-pair correlations to create entanglement or avoid entanglement for high purity. Regardless, characterization of the photon-pair source is an indispensable step towards defining the correlations present in the photon pairs so as to utilize the source for appropriate applications.

There are four main DOFs when it comes to characterizing photon pairs: polarization \cite{Fan_J2007,Zhou2013,Daniel2021}, temporal \cite{Kuzucu2008,Dong2014,Aguayo2022}, spectral \cite{Wasilewski:06,Lugani:20}, and transverse spatial degrees of freedom \cite{Cruz-Delgado2014,Pourbeyram2016}. For polarization and temporal degrees of freedom, there are many well-developed methods that can measure the polarization and temporal mode of photons in a relatively straightforward way by using linear optics, single-photon detectors, and coincidence electronics to reveal the correlations. Thus, here, we will focus more on the characterization methods for two other degrees of freedom -- spectral and transverse spatial modes -- that may require a more careful approach.

Before we delve into describing characterization methods for specific degrees of freedom, we introduce metrics that describe the degree of quantum state purity of the source across all degrees of freedom: quantum state purity $P$, the second-order autocorrelation function $g^{(2)}$, and the Hong-Ou-Mandel dip. Using these quantities, we can assess whether a fiber-based photon-pair source has high enough quantum state purity for employment in diverse quantum information processing applications.

\emph{Quantum State Purity $P$} For a quantum state defined by the density operator $\rho$, the purity is generally defined as in Eq.~\ref{eq:purityRho},
\begin{align} 
    P&\equiv\mathrm{Tr}\left[\rho^2\right], \label{eq:purityRho}
\end{align}
where $P=1$ for a pure state and $P=1/d$ for a $d$-dimensional maximally mixed state \cite{nielsen_chuang_2010}. For photon-pair generation, $P$ represents the quantum state purity of the heralded single-photon state (sometimes called single-photon purity) when one of the two daughter photons in a photon-pair source is used to indicate the presence of another. The heralded photon's quantum state purity $P$ is inversely proportional to the effective number of eigenmodes, described by Schmidt number $K$
\begin{align} 
    P = \frac{1}{K} \label{eq:puritySchmidt}.
\end{align}
Given that the signal-idler photon-pair state can be written in orthonormal states of signal and idler using Schmidt (singular value) decomposition as $f_{s,i}=\sum_j \sqrt{\lambda_j}g_{s,j}g_{i,j}$, the Schmidt number is defined as,
\begin{equation}
    K=\frac{1}{\sum_j\lambda_j^2},
\end{equation}
where the Schmidt coefficients $\lambda_j$ satisfy the normalization condition, $\sum_j\lambda_j=1$. Qualitatively speaking, $K$ describes the effective number of populated eigen-Schmidt modes or the degree of entanglement.

\emph{Second-order Autocorrelation function $g^{(2)}$}
 \label{g2Measurement}
There are various types of second-order correlation measurements that are useful in characterizing photon-pair sources, e.g., to determine signal-idler cross-correlations and the degree to which the source is producing true single-photon states; for a discussion of these we refer the reader to~\cite{spring2013chip}. Here we focus on the \textit{autocorrelation}  $g^{(2)}$, defined as
\begin{equation} \label{eq:g2Func}
    g^{(2)}=g^{(2)}(\tau=0)=\left.\frac{\langle n_1(t)n_2(t+\tau)\rangle}{\langle n_1(t)\rangle\langle n_2(t)\rangle}\right\rvert_{\tau=0}=\frac{N_\mathrm{coinc}R}{N_1N_2},
\end{equation}
which can be used to determine the quantum state purity through the use of a beam splitter in just one arm of the photon-pair source and measuring the two-fold coincidences at the outputs~\cite{zielnicki2015engineering,zielnicki2018joint}. This type of $g^{(2)}$ measurement can be simply described as a Hanbury Brown-Twiss-type of interferometric measurement conducted on either signal or idler side of the photon-pair source. For a photon-pair source with $K$ thermal eigenmodes (Schmidt number), the following relation holds~\cite{christ2011probing,blauensteiner2009photon}:
\begin{equation} \label{eq:g2}
    g^{(2)}=1+\frac{1}{K}=1+P.
\end{equation}
The more thermal eigenmodes the quantum state is comprised of, the more Poissonian ($\langle n^2\rangle=\Bar{n}$ where $\Bar{n}$ is the average photon number), and less thermal or super-Poissonian ($\langle n^2\rangle>\Bar{n}$), it is~\cite{mauerer2009colors}. If in addition the photons are heralded, the three-fold coincidences of two signal (idler) ports and one idler (signal) port $g^{(2)}_{s1,s2,i}$ ($g^{(2)}_{s,i1,i2}$) gives 0 when the heralded single-photon state is pure, and greater than 0 when it has multiple modes~\cite{Smith:09,feng2019chip}. Temporal resolution of the $g^{(2)}$ measurement is dependent on the relative magnitude of detector response time compared to the coherence time of the heralded single photon. When the detector response time is longer than the coherence time,  the time-integrated version of the correlation function $g^{(2)}$ should be considered (see \cite{zielnicki2018joint} for more detail). In addition, this measurement is conditioned upon the assumption that there are a negligible number of photons arriving during the APD's dead-time.

\emph{Hong-Ou-Mandel dip}
 Hong-Ou-Mandel (HOM) interferometry is a four-fold coincidence measurement involving two photon pairs originating from two separate photon-pair sources~\cite{Soller2011}. For sources with heralded single-photon purity $P$, the HOM interference exhibits a dip whose visibility $V$  satisfies $V=P$. Two-source HOM interferometry thus provides a direct measurement of quantum state purity, but it suffers from long acquisition times that can extend up to days~\cite{zielnicki2015engineering,zielnicki2018joint}. The purity derived from the autocorrelation $g^{(2)}$ measurement provides an upper limit to the purity acquired from the two-source HOM measurement since HOM relies on the two different sources having the same purity.

\subsection{Spectral characterization} \label{spectralCharact}

\begin{figure*}[t!]
\begin{center}
    \includegraphics[width=0.9\linewidth]{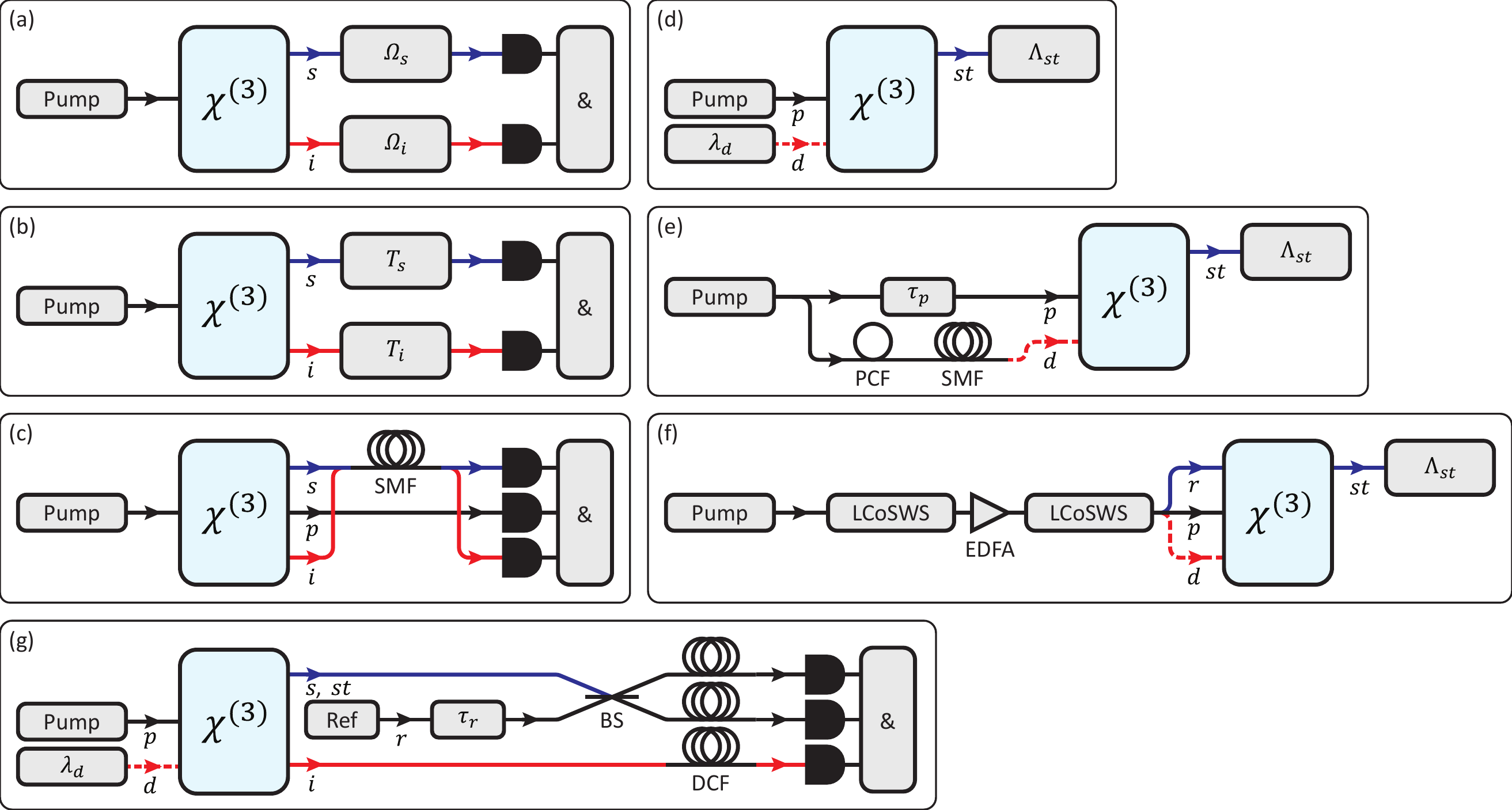}
          \caption{Spectral characterization methods for SFWM photon-pair source. The subscripts represent $p$: pump,  $s$: signal, $i$: idler, $d$: seed, $st$: stimulated signal, $h$: herald, and $r$: reference pulse. (a) Scanning monochromator measurement; $\Omega_{s,i}$: monochromators for signal and idler photons. (b) Fourier transform spectroscopy; $T_{s,i}$: Michelson or common path interferometers to change the relative temporal delay in each path. (c) Dispersive fiber spectroscopy; a single mode fiber (SMF) for applying dispersion on both signal and idler, temporally spreading spectral components. Single-photon detectors are connected to a coincidence counter ($\&$). (d) Stimulated emission tomography; a tunable seed laser ($\lambda_d$) and a spectrometer ($\Lambda_{st}$) for the stimulated signal. (e) Chirped supercontinuum seed measurement; temporal delay $\tau_p$, photonic crystal fiber (PCF) for supercontinuum generation, and a SMF for chirping. (f) Phase sensitive tomography; one pump laser is combined with liquid-crystal-on-silicon wave shapers (LCoSWS) and an erbium-doped fiber amplifier (EDFA) to produce pump, seed, and broadband phase reference pulses. (g) Intensity interferometry; signal is interfered on a beam splitter (BS) with a weak reference pulse (Ref) that has a relative temporal delay $\tau_r$. Dispersion compensated fibers (DCF) are used for the same purpose as the SMF in (c) dispersive fiber spectroscopy.}
    \label{fig:SpectralCharact}
\end{center}
\end{figure*}

While the above metrics provide information on overall quantum state purity, it is often useful to characterize in specific degrees of freedom to address poor fidelity to the target state. There are a number of different strategies to measure the spectral correlations of photon pairs, including making direct joint measurements of signal and idler spectral components, performing measurements based on interference, stimulating the FWM process, and combinations of these techniques (see Fig.~\ref{fig:SpectralCharact}). When measuring the joint spectral function, it is sometimes valid to assume that the joint spectral phase is flat over the spectrum and thus an intensity measurement of the joint spectrum is sufficient to characterize the correlations. Other methods can be realized that provide the full, complex JSA, $K$~\cite{jizan2016phase}, but may be susceptible to artifacts from, e.g., temporal chirp in the pump. In both cases, the JSI gives a lower bound on the Schmidt number $\tilde{K}$ (or upper bound on the purity). The derived purity can be compared to the heralded single-photon purity $P$ obtained from autocorrelation $g^{(2)}$ measurements (Eq.~\ref{eq:g2}), which includes information on the joint spectral phase. Exploiting this property, inadequacies present in joint spectral measurements can be revealed through comparison with the purity obtained from autocorrelation $g^{(2)}$ measurements~\cite{zielnicki2015engineering,Blay2017,Lugani:20,Paesani2020}.

\subsubsection{Scanning monochromator measurement} \label{scanningMonochromator}
The scanning monochromator measurement (Fig.~\ref{fig:SpectralCharact} (a)) is the most direct method of measuring the spectral correlations of photon pairs. It involves two-dimensional raster scanning of the signal and idler photon frequencies by rotating the gratings in two monochromators placed in each of the signal and idler paths, and counting coincidences with single-photon detectors at individual grating positions~\cite{zielnicki2018joint,zielnicki2015engineering}. The 2D scan is processed to provide the joint spectral intensity (JSI). Although it may seem quite straightforward to implement, this method has a major drawback of requiring a long data acquisition time (on the order of many hours) in order to achieve a reasonable resolution, due to the monochromator filtering a very narrow spectral window, leading to relatively low pair collection efficiencies ($\sim$ 0.1 \%). Thus, compared to other measurement schemes, JSI measurement through the scanning monochromator technique suffers from low SNR. 

\subsubsection{Fourier transform spectroscopy} \label{FTspectroscopy}
In Fourier transform spectroscopy (Fig.~\ref{fig:SpectralCharact} (b)), temporal interference between either single photons or photon pairs are measured and then Fourier transformed back to the spectral domain using an interferometer with variable temporal delay in one arm. If only one photon from the pair is sent to the interferometer, i.e., one-dimensional Fourier transform spectroscopy, the single photon spectrum is measured. Sending both photons through the interferometer allows for measuring the JSI, i.e., two-dimensional Fourier transform spectroscopy~\cite{wasilewski2006joint}. Implementations include Michelson or common path polarization-based interferometer that uses a pair of quartz wedges~\cite{zielnicki2015engineering}. The time-dependent intensity signal $I(\tau)$ is Fourier transformed back to a frequency-dependent spectrum $I(\omega)$ for one dimension as
\begin{equation}
    I(\omega)\propto\int^\infty_0d\tau\left(I(\tau)-\frac{1}{2}I(\tau=0)\right)\cos(\omega\tau),
\end{equation}
and for two dimensions, with appropriate filtering, as
\begin{equation}
\begin{split}
    I(\omega_s>0,\omega_i>0)&=\int d\tau_s d\tau_i\,\tilde{I}(\tau_s,\tau_i)\exp(i\omega_s\tau_s+i\omega_i\tau_i)\\
    &\propto\delta(\omega_s)\delta(\omega_i)\langle\hat{N}_s\hat{N}_i\rangle+\frac{1}{2}\delta(\omega_s)\langle\hat{N}_s\hat{I}_i(\omega_i)\rangle\\
    &\quad+\frac{1}{2}\delta(\omega_i)\langle\hat{N}_i\hat{I}_s(\omega_s)\rangle_+\frac{1}{4}I(\omega_s,\omega_i)\\
    &\propto(\textrm{total coincidence counts at origin})\\
    &\quad+(\textrm{heralded idler spectrum at }\omega_i\textrm{ axis})\\&\quad+(\textrm{heralded signal spectrum at }\omega_s\textrm{ axis})\\
    &\quad+(\textrm{JSI}),
\end{split}
\end{equation}
 where $\hat{I}_{s,i}(\omega_{s,i})$ and $\hat{N}_{s,i}=\int d\omega\hat{I}_{s,i}(\omega)$ are operators that each represents the spectral intensity and total number of signal or idler photons. Although two-dimensional Fourier spectroscopy requires $N^2$ data points compared to $N$ for the one-dimensional case, this can be reduced to $N$ by performing diagonal Fourier spectroscopy if it can be assumed that the JSI is approximately a Gaussian ellipse aligned along the diagonal ($\omega_s+\omega_i$) and anti-diagonal axes ($\omega_s-\omega_i$). Then, a 1D scan along the $t_s+t_i$ axis -- the $\omega_s+\omega_i$ axis in Fourier space -- gives diagonal ($\sigma_d$) and anti-diagonal bandwidth ($\sigma_a$) information of the JSI. The ratio of the two $r\equiv\sigma_d^2/\sigma_a^2$ can then be utilized to calculate the quantum state purity.

\subsubsection{Dispersive fiber spectroscopy} \label{dispersiveFiberSpectroscopy}
Dispersive fiber spectroscopy (Fig.~\ref{fig:SpectralCharact} (c)) uses frequency-dependent photon time-of-arrival information to measure the spectral correlation of photon pairs. As an example, in~\cite{zielnicki2015engineering,zielnicki2018joint}, both the 810 nm signal and idler photons are sent into the same 400 m-long fiber that introduces a dispersion of $-120$ ps/nm/km, or group velocity dispersion of $+41.8 \mathrm{\ fs}^2$/mm. Then, the photon pairs are counted in coincidences triggered by the detection of pump photons, and the 2D JSI is recreated based on the following relation: $\Delta\lambda_{s,i}\approx c(t_{s,i}-t_p)$ where $t_{s,i}$ and $t_p$ correspond to timestamps of the signal, idler, and pump (sync trigger). Although in this particular example the photon-pairs were created from a Type-I degenerate spontaneous parametric down-conversion (SPDC) process in $\beta$-BBO (barium borate) crystal, the measurement technique can be equivalently applied to fiber-based SFWM sources, as can be seen in the intensity interferometry example discussed in subsection~\ref{intensityInterferometry}.

The resolution of the measurement can be improved by higher dispersion or longer propagation distance in the media; but at the same time can be limited by the photon loss and timing jitter of the detection. There exists an algorithmic method to deconvolve the effect of timing jitter but it is limited to the case where the jitter is on the order of the narrow bandwidth of the JSI ellipse. In regards to improving the loss, use of a chirped fiber Bragg grating (CFBG) was proposed~\cite{davis2017pulsed}. As compared to ordinary single mode fiber to introduce group delay dispersion, a CFBG combined with a fiber circulator can provide a relatively low-loss all-fiber solution (faster data acquisition by up to a factor of 20).

\subsubsection{Stimulating the process with a seed: stimulated emission tomography} \label{SET}
First theoretically proposed in~\cite{liscidini2013stimulated}, and experimentally demonstrated in fiber in~\cite{fang2014fast}, stimulated emission tomography (SET) is a relatively fast (order of tens of minutes acquisition time) measurement technique that can yield high-SNR JSI results (Fig.~\ref{fig:SpectralCharact} (d)). Simply put, SET relies on having a tunable narrowband seed laser at either the signal or idler wavelength to stimulate the FWM processes. Then the number of stimulated photons is proportional to the product of the number of corresponding spontaneous photon pairs and the stimulating seed photon flux. 

The SNR of the JSI obtained from SET is high compared to the other JSI characterization techniques described above even when normalized by the acquisition time and spectral resolution to remove the any equipment-dependent bias (see Table. 3 in~\cite{zielnicki2018joint}). Thus, SET can be applied to efficiently compare spectral correlations for varying fiber lengths~\cite{fang2014fast} or characterize an entangled photon-pair source in multiple dimensions, such as in both polarization and frequency degrees of freedom~\cite{fang2016multidimensional}.

\subsubsection{Chirped supercontinuum seed} \label{chirpedSupercontinuumSeed}
The spectral characterization schemes described above can be combined to provide faster measurement and/or sensitivity to the joint spectral phase of the photon pair. The chirped supercontinuum seed scheme (Fig.~\ref{fig:SpectralCharact} (e) combines SET with a highly-dispersive fiber to achieve even faster acquisition times (5-30 seconds depending on the resolution)~\cite{erskine2018real}. In contrary to the original SET demonstration that requires a separate narrowband tunable continuous-wave (CW) seed laser, this scheme uses a single pulsed laser for both pump and seed. Forming a Mach-Zehnder-like interferometer with the pulsed laser at the input, the beam in one arm is sent into a photonic crystal fiber (PCF) for supercontinuum generation (broadband white light); by sending the white light into a 8 m-long single mode fiber to introduce chirp, the frequency components of the white light are spread in the temporal domain to result in a chirped supercontinuum. With this chirped supercontinuum seed in hand, by adjusting the temporal delay of the temporally narrow-band pump beam,  the seed wavelength can be effectively tuned. The seed wavelength calibration as a function of pump temporal delay can be performed by utilizing nonlinear interaction between the pump and seed that results in a $\sim1 \%$ dip in the residual seed spectrum.

 This chirped supercontinuum seed measurement can be limited by the spectrometer and pump temporal delay resolution, and ultimately by the pulse width of the pump source -- it requires a short pump pulse ($\lesssim 10 ps$) to prevent the finite temporally-selected seed bandwidth from broadening the JSI. Nevertheless, this method can be used to compare JSI correlations induced by varying pump bandwidth.

\subsubsection{Phase-sensitive tomography} \label{phaseSensitiveTomography}
Phase-sensitive tomography measurement (Fig.~\ref{fig:SpectralCharact} (f)) combines SET with phase-sensitive amplification (PSA) to unveil both the joint spectral amplitude (JSA) and joint spectral phase (JSP) of the photon-pair source, as shown for integrated silicon nanowire and a highly nonlinear fiber source in Ref.~\cite{jizan2016phase}. A single broadband laser at 1555 nm with 30 nm bandwidth is combined with two dynamically tunable filters [liquid-crystal-on-silicon wave shaper (LCoSWS)] and an erbium-doped fiber amplifier (EDFA) to produce three pulses: pump, seed, and broadband phase reference pulse. These three are then sent into the photon-pair source, whose output light is then analyzed with a optical spectrum analyzer for interference. The interference, especially in the phase reference spectrum, is the consequence of four different phase settings applied on the seed pulse. By taking clever linear combinations of the four measurements, the full JSA can be obtained.

Though the phase-sensitive tomography method's resolution can be restricted by the finite bandwidth of the seed that can be created with the LCoSWS (10 GHz in this case), the resolution has been sufficient to reveal otherwise hidden features originating from both intentional (pump chirp introduced by the second LCoSWS) and unintentional (possible dispersive effects from EDFA) phase correlations in the system. It is also sensitive to small spectral oscillations not observable in the JSI. 

\subsubsection{Intensity interferometry} \label{intensityInterferometry}
The recently developed intensity interferometry (Fig.~\ref{fig:SpectralCharact} (g)) method  in~\cite{thekkadath2022measuring} couples SET, dispersive fiber spectroscopy, and Fourier transformation to measure both the JSA and JSP of photon pairs. Although the method may seem comparable to phase sensitive tomography (subsection~\ref{phaseSensitiveTomography}), it does not require additional phase stability, nonlinearity, or spectral shaping. And unlike Fourier transform spectroscopy (subsection~\ref{FTspectroscopy}), the intensity interferometry method does not scan a relative time delay, but rather relies on a fixed delay. 

To perform the measurement, signal photons are interfered with a weak reference pulse on a beam splitter and spectral intensity correlations are measured at two output ports ($1,2$). Dispersion-compensating fibers (dispersion of $-997 \mathrm{\ ps/nm}$) are implemented at the output ports to map frequency to arrival time, and a time tagger is used to build a coincidence histogram. The reference pulse is temporally delayed by a fixed amount $\tau$ and attenuated to the single-photon level using neutral-density (ND) filters. The measured 2D spectral intensity $\langle\hat{G}(\omega_1,\omega_2)$ is then Fourier transformed and inverse Fourier transformed with appropriate Fourier filtering to obtain the JSA ($f(\omega_1,\omega_2)$) and JSP ($\mathrm{arg}\{f(\omega_1,\omega_2)\}$) of the signal.

This technique's resolution can be limited if the reference pulse bandwidth is greater than that of the signal. It also requires \emph{a priori} measurement of the reference pulse spectral mode, using classical pulse characterization techniques, e.g., FROG and SPIDER. In addition, in the case of a photon-pair source, it requires measuring 3-fold coincidences ($N(\omega_1,\omega_2,\omega_h)$) of signal, herald (idler), and the reference photons, which leads to a few hours of acquisition time. However, the process can be accelerated by using a tunable CW laser as a seed to perform SET by measuring 2-fold coincidences between stimulated signal and reference, yielding ($N(\omega_1,\omega_2,\omega_{seed})$) instead. The method can also be simplified by measuring 2-fold coincidences directly, without heralding or seeding. In both cases, although fringe visibility is reduced, employing adequate Fourier filtering makes the method relatively resilient to these alterations.

The method was also experimentally compared to other methods discussed above. In comparing the uncorrelated JSA produced from spectrally filtering the pump with a phase-correlated JSA generated from chirping the pump with a 5 m-long single mode fiber, the intensity interferometry method was able to uncover the correlations arising from the joint spectral phase: $K=1.04\ (g^{(2)}=1.96)\rightarrow K=1.48\ (g^{(2)}=1.68)$.

\subsection{Transverse spatial mode characterization}
For characterizing the transverse spatial correlations of photon pairs, parallels can be drawn with the strategies in Section~\ref{spectralCharact} for spectral characterization. For spatial characterization, there exist analogues of direct transverse intensity scanning measurement, measurements based on interference, and stimulated FWM processes. In contrast to spectral characterization, however, there are not as many examples due to the infancy of the field in spatial degrees of freedom. Yet, we can imagine employing similar, relatively well-developed techniques used to characterize free-space SPDC sources~\cite{hiekkamaki2021high}, integrated waveguide sources~\cite{feng2019chip}, or even classical sources~\cite{carpenter2012degenerate,carpenter2016complete,fontaine2019laguerre}. These strategies can be used to reveal and verify transverse spatial mode correlations and entanglement present in photon pairs created from optical fiber, e.g., by quantum state tomography (QST)~\cite{kim2022transverse}.

\subsubsection{Transverse intensity scanning}
Analogous to rasterized scanning of monochromators in subsection~\ref{scanningMonochromator}, transverse scanning of spatial intensity can be employed to characterize the spatial correlations~\cite{Cruz-Delgado2016}. In Ref.~\cite{Cruz-Delgado2016}, since frequency-transverse mode hybrid correlations were explored, both the spatial scanning and the spectral monochromator measurement were conducted simultaneously on the arms of signal and idler, e.g., spatial measurement on signal heralded by the spectral measurement on idler. However, for general purposes, the output collimator lens on the fiber-based photon-pair source can be transversely translated in $x, y$ directions to record the corresponding coincidences of signal and idler using multi-mode fibers (MMF) attached to single-photon detectors like avalanche photodiodes (APD).

Naturally, though, this measurement technique can suffer from long acquisition times depending on the resolution required. Thus, for higher-order transverse modes that may have small detailed features, this may not be the most optimal measurement scheme. However, it is indeed one of the most direct methods to record the transverse spatial intensity correlations of photon-pairs owing to the simplicity of the experimental setup.

\subsubsection{Transverse mode projection measurement}
An alternative way of characterizing the transverse spatial correlations of photon-pairs is to project individual photons into different transverse mode basis states, thus mapping 2D-transverse intensity into probability amplitude (intensity) of projection measurements~\cite{kim2022transverse,hiekkamaki2021high}. In classical communication protocols, this method is sometimes referred to as mode sorting or mode demultiplexing~\cite{carpenter2012degenerate,carpenter2016complete,fontaine2019laguerre}. The main principle is to use a phase-only spatial light modulator (SLM) to impose different transverse spatial phases on the incident photons and couple the output light into a single-mode fiber (SMF). If the applied phases is conjugate to that of the incident photons, the SLM-reflected light will have an approximately fundamental Gaussian mode shape at the far field, in this case at the input of the SMF. On the other hand, if the SLM phase mask is different from the phase of the photon, only a portion of the intensity proportional to the quantum projection amplitude will be coupled into the SMF.

This method may require careful engineering of the SLM phase masks and some \textit{a priori} knowledge of the transverse modes of the photon-pairs to accurately define the basis states that best represent the transverse modes of the photons. Despite these possible experimental challenges, transverse mode projection is still one of the most versatile methods in the sense that can measure the transverse state of the photons in quantum superposition and thus can be used for quantum state tomography (QST)~\cite{altepeter2005photonic}, which requires measuring in different basis states and superposition states thereof~\cite{kim2022transverse}. In turn, QST is then able to statistically reconstruct the density matrix $\rho$ of the quantum photon-pair state and be subsequently analyzed for any transverse spatial entanglement present in the system as well as other quantum metrics such as quantum state purity.

\subsubsection{Temporal correlation measurement}
Spatial correlation present in the photon pairs can be indirectly observed by transverse mode-dependent time of arrival, as was shown in a proof-of-principle type of experiment in Ref.~\cite{sulimany2022all}. This technique resembles dispersive fiber spectroscopy introduced in subsection~\ref{dispersiveFiberSpectroscopy}, especially in that it also exploits the group delay dispersion of a SMF to map the frequency of signal and idler photons into the temporal DOF. Beyond the fact this method shares the same types of limitations as dispersive fiber spectroscopy, like timing jitter and fiber-dependent loss, this technique has an additional critical flaw when it comes to using it to assessing the quantum nature of the given spatial correlations, e.g., entanglement. Because it only relies on temporal correlation in intensity and not phase, it is unable to distinguish quantum entangled states from classically mixed states, and thus cannot verify the exact quantum state of photon pairs to determine suitability for quantum applications.

\subsubsection{Stimulated-emission-based transverse mode measurement}
Since spatial correlations residing in the photon pairs are the result of transverse phasematching conditions of the FWM interaction in optical fiber, stimulated-emission-based measurement can facilitate the characterization of spatial correlations. As was explored in \cite{kim2020stimulated,kim2020towards}, not only can transverse mode-dependent spectral correlations be observed, but also direct imaging of stimulated signal photons can be performed without using single-photon level detectors that require long exposure times. 

Because the transverse modes are often nondegenerate in frequency, they exhibit hybrid correlations, as was studied in Ref.~\cite{Cruz-Delgado2016}. Thus, changes in the spatial correlation of the signal and idler photons can be observed via spectral correlation changes in the JSI~\cite{liu2022engineering,liu2022shaping} (the orbital angular momentum (OAM) mode in fiber was used in these references). As was seen in the evolution of spectral characterization techniques using SET, there lies an exciting path forward for transverse spatial characterization utilizing the stimulated process approach.


\section{Conclusions}

In this tutorial paper we have presented a broad overview of the topic of photon-pair generation through the spontaneous four wave mixing (SFWM) process in optical fibers.    This overview covers the SFWM two-photon state and the various possible phasematching configurations, as well methods for photon-pair factorability, frequency tunability, and bandwidth control.   It covers the use of dual pumps,  in the form of, both, frequency non-degenerate pumps and counter-propagating pumps.   It also covers the generation of entanglement in various degrees of freedom or combinations of degrees of freedom, including energy-time, polarization, frequency-transverse mode, frequency-polarization, and discrete frequency. Finally, this paper covers characterization techniques for photon pairs generated by SFWM in optical fibers, in both the spectral and spatial domains.  It is hoped that this tutorial paper will constitute a valuable resource for researchers who may be new in the area, contributing to the consolidation of fiber-based implementations of photon-pair sources.



\bibliography{TutorialReferences,characterizationReferences,ext_literature}

\end{document}